\documentclass[11pt,a4paper,aps]{article}
\usepackage{etex}
\usepackage{inputenc}
\usepackage[T1]{fontenc}
\usepackage{lmodern}					% harmonie des polices
\usepackage{textcomp}				% table de symboles
\usepackage{verbatim}			% pour faire des commentaires par block avec begin{comment}...end{comment}
\usepackage[english]{babel}
\usepackage[toc,page]{appendix} 		% pour faire des annexes
\usepackage{cite}
\usepackage{amsmath,					% packages mathmatiques
			amsthm,
			amsfonts,
			amssymb, 
			mathtools,
			mathrsfs}
\usepackage{enumitem} 
\usepackage{todonotes}
\usepackage{stmaryrd}

\usepackage[all]{xy}
\usepackage{tikz-cd}

\DeclareMathAlphabet{\mathpzc}{OT1}{pzc}{m}{it}

\usetikzlibrary{tikzmark}
\usetikzlibrary{arrows}
\newcounter{diagram}

\usepackage{graphicx}				% pour inclure des figures (.pdf, .jpg)
\usepackage{tabularx}				% pour inclure des tableaux (careful)

\usepackage{pict2e}

\makeatletter
\DeclareRobustCommand{\loplus}{\mathbin{\mathpalette\dog@lsemi{+}}}
\DeclareRobustCommand{\lotimes}{\mathbin{\mathpalette\dog@lsemi{\times}}}
\DeclareRobustCommand{\roplus}{\mathbin{\mathpalette\dog@rsemi{+}}}
\DeclareRobustCommand{\rotimes}{\mathbin{\mathpalette\dog@rsemi{\times}}}

\newcommand{\dog@rsemi}[2]{\dog@semi{#1}{#2}{-90,90}}
\newcommand{\dog@lsemi}[2]{\dog@semi{#1}{#2}{270,90}}
\newcommand{\dog@semi}[3]{%
  \begingroup
  \sbox\z@{$\m@th#1#2$}%
  \setlength{\unitlength}{\dimexpr\ht\z@+\dp\z@\relax}%
  \makebox[\wd\z@]{\raisebox{-\dp\z@}{%
    \begin{picture}(1,1)
    \linethickness{\variable@rule{#1}}
    \roundcap
    \put(0.4,0.425){\makebox(0,0){\raisebox{\dp\z@}{$\m@th#1#2$}}}
    \put(0.4,0.4){\arc[#3]{0.5}}
    \end{picture}%
  }}%
  \endgroup
}
\newcommand{\variable@rule}[1]{%
  \fontdimen8  
  \ifx#1\displaystyle\textfont3\else
    \ifx#1\textstyle\textfont3\else
      \ifx#1\scriptstyle\scriptfont3\else
        \scriptscriptfont3\relax
  \fi\fi\fi
}
\makeatother

\usepackage{authblk}   % pour les auteurs

\definecolor{pur}{RGB}{255,102,102}

\usepackage[hmargin=2.2cm,vmargin=2.5cm]{geometry}		% adapter les dimensions de la page
\numberwithin{equation}{section}

\newtheorem*{definition*}{Definition}

\newtheorem*{theoreme*}{Theorem (Getzler \cite{Getzler:2010})}

\newtheorem*{lemme*}{Lemma}

\newtheorem*{corollaire*}{Corollary}

\theoremstyle{remark}

\usepackage[colorlinks]{hyperref}			% liens dans le .pdf

\newcommand{\bigslant}[2]{{\left.\raisebox{.2em}{$#1$}\middle/\raisebox{-.2em}{$#2$}\right.}}

\renewcommand{\leq}{\leqslant}
\renewcommand{\geq}{\geqslant}

\title{Infinity-enhancing of Leibniz algebras}

\author[1]{Sylvain Lavau\thanks{lavau@math.univ-lyon1.fr (corresponding author)}}
\author[2,3]{Jakob Palmkvist\thanks{jakob.palmkvist@oru.se}}
\affil[1]{\small \emph{IMJ-PRG, Universit\'e Paris Diderot, Paris, France.}}
\affil[2]{\small \emph{School of Science and Technology, \"Orebro University, \"Orebro, Sweden.}}
\affil[3]{\small \emph{Mathematical Sciences, Chalmers University of Technology and University of Gothenburg,\authorcr \it G\"oteborg, Sweden.}}
\date{}

\begin{document}

\maketitle

\vspace{-.5cm}

\begin{center}
\today
\end{center}

\abstract{
%\SL{I added a few words in the sentence below because a dgLa is a $L_\infty$-algebra so I thought people could miss the importance of this sentence. Please adapt following your impressions.}

We establish a correspondence between infinity-enhanced Leibniz algebras, recently introduced 
%\cite{Bonezzi:2019ygf} 
in order to encode tensor hierarchies \cite{Bonezzi:2019ygf}, and 
differential graded Lie algebras, which have been already used in this context.
We explain how any Leibniz algebra gives rise to a differential graded Lie algebra with a corresponding infinity-enhanced Leibniz algebra.
%show that the recent formulation of tensor hierarchies in terms of
%\emph{infinity-enhanced Leibniz algebras}
%\cite{Bonezzi:2019ygf}
%is a consequence of the differential graded Lie algebra structure already used in this context by
%constructions from a Leibniz algebra or a tensor hierarchy algebra.
Moreover, by a theorem of Getzler, this differential graded Lie algebra canonically induces an $L_\infty$-algebra structure on the suspension of the underlying chain complex.
We explicitly give the brackets to all orders and 
show that they 
agree with the partial results
obtained from
the infinity-enhanced Leibniz algebras in \cite{Bonezzi:2019ygf}.
}

\tableofcontents

\section{Introduction}

Leibniz algebras (also known as Loday algebras or Leibniz-Loday algebras \cite{Loday}) have during the last years attracted attention for their applications to gauge theories where the gauge
variation $\delta_{x}y$
of one parameter $y$ with respect to another $x$ is not antisymmetric under the interchange of $x$ and $y$ (but where the symmetrisation leads to a parameter that acts trivially on the fields). Such situations occur for example
in the embedding tensor formulation of
gauged supergravity \cite{deWit:2002vt,deWit:2004nw,deWit:2005hv,deWit:2008ta,Trigiante:2016mnt} as well as in extended geometry
\cite{%Hull:2009mi,
Hull:2009zb,Coimbra:2011ky,Berman:2012vc,Cederwall:2013naa,Cederwall:2013oaa,Aldazabal:2013mya,Hohm:2013nja,Hohm:2013pua,Hohm:2013vpa,Hohm:2013uia,Hohm:2014fxa,Hohm:2015xna,Abzalov:2015ega,Wang:2015hca,Cederwall:2015ica,Musaev:2015ces,%Palmkvist:2015dea,
Berman:2015rcc,Deser:2016qkw,Bossard:2017aae,Cederwall:2017fjm,%Deser:2018oyg,
Cagnacci2019,Cederwall:2018aab,Hohm:2018ybo,Bossard:2018utw,Hohm:2019wql}, and give rise to a tensor hierarchy of gauge parameters,
potentials and field strengths.

In \cite{Bonezzi:2019ygf} it was shown that the general gauge theory based on any Leibniz algebra 
leads to an extension of it to an {\it infinity-enhanced Leibniz algebra}, and 
it was proposed that this is the most general algebraic structure that enables the construction of the full tensor hierarchy. It is a further development of
the notion of an {\it enhanced Leibniz algebra}, originally introduced in \cite{Strobl:2016aph, Strobl:2019hha}, to encode general gauge-invariant action functionals for coupled 1- and 2-form gauge fields with kinetic terms. 
In this paper we will show that any Leibniz algebra can be
`infinity-enhanced' in the sense that it canonically gives rise to a differential graded Lie algebra concentrated in non-negative degrees,
and that any such differential graded Lie algebra structure in turn implies the axioms of an infinity-enhanced Leibniz algebra. This settles the problem of existence of such a structure, and provides an alternative algebraic structure encoding the tensor hierarchy that seems much simpler.

The idea of encoding the tensor hierarchy by a differential graded Lie algebra is not new.
It was originally proposed in \cite{Palmkvist:2013vya, Greitz:2013pua}, and was based on the relation between
tensor hierarchies and certain Borcherds-Kac-Moody superalgebras \cite{Henneaux:2010ys,Palmkvist:2011vz,Cederwall:2015oua,Palmkvist:2015dea,Cederwall:2018aab}.
The construction of such
a structure
from a Leibniz algebra $V$, together with a Lie algebra $\mathfrak{g}$ that acts on it and
an embedding tensor $\Theta : V \to \mathfrak{g}$ satisfying certain compatibility conditions was given in \cite{Lavau:2017tvi}\footnote{%
Unfortunately, the differential graded Lie algebra in \cite{Lavau:2017tvi}
was also called `tensor hierarchy algebra', but there is an important difference: the tensor hierarchy algebra in \cite{Palmkvist:2013vya} is 
{\it a priori} not a differential graded Lie algebra, but a $\mathbb{Z}$-graded Lie superalgebra with a subspace at degree $-1$
accommodating all
possible embedding tensors satisfying the representation constraint.
%in a certain representation of $\mathfrak{g}$. 
Restricting it to a one-dimensional subspace spanned by one particular embedding tensor leads to the
differential graded Lie algebra called `tensor hierarchy algebra' in \cite{Lavau:2017tvi}.
}.
In the present paper, we will show that for any Leibniz algebra $V$ there is a canonical choice of Lie algebra $\mathfrak{g}$, namely $\mathfrak{gl}(V)$, and an embedding tensor $\Theta : V \to \mathfrak{g}$
satisfying these compatibility conditions. With this choice, the construction in \cite{Lavau:2017tvi} leads to the extension of any Leibniz algebra
to a differential graded Lie algebra concentrated in non-negative degrees, without explicit reference to a Lie algebra or an
embedding tensor as initial data.  
%\SL{this part is sloppy}
%\textbf{As already mentioned, this differential graded Lie algebra (like any other differential graded Lie algebra concentrated in non-negative degrees)
As we will show, the axioms defining an infinity-enhanced Leibniz algebra follow from this differential graded Lie algebra structure.
%(like any other differential graded Lie algebra concentrated in non-negative degrees).
Conversely, we will also show that any infinity-enhanced Leibniz algebra
gives rise to a differential graded Lie algebra concentrated in non-negative degrees by the canonical choice of Lie algebra $\mathfrak{g}$ and 
embedding tensor $\Theta : V \to \mathfrak{g}$. 
This correspondence %\footnote{One can show that under particular conditions on the differential graded Lie algebra, this correspondence can be made one-to-one. We will not address this topic in this paper since it would require additional mathematical details that would increase the length of the present paper and bring unnecessary confusion, with respect to the goal of the paper.} 
encourages us to think that the most natural and useful algebraic structure encoding the tensor hierarchy is the differential graded Lie algebra defined by the Leibniz algebra~$V$.

Due to the fact that the skew-symmetric part of the Leibniz product is not a Lie bracket (it does not satisfy the Jacobi identity), the gauge theory based on a Leibniz algebra $V$ does not behave as the usual Yang-Mills gauge theory. Rather, covariance of the field strengths is guaranteed at the cost of adding higher fields, which leads to the famous tensor hierarchy. %of higher gauge parameters, potentials and field strengths.
The algebraic structure that the gauge parameters form turns out to be a (non-negatively graded)
{\it $L_\infty$-algebra}, rather than a Lie algebra. An $L_\infty$-algebra is a generalization of a differential graded Lie algebra, where the Jacobi identity is weakened to be satisfied only \emph{up to homotopy}. This implies that higher brackets have to be introduced, satisfying
`higher Jacobi identities'. The number of brackets can be infinite if the chain complex hosting the $L_\infty$-algebra structure is not bounded (see \cite{Lada:1992wc, Marklada} for an introduction to this topic, and
\cite{Palmer:2013pka,Lavau:2014iva,
Ritter:2015ymv,Saemann:2017rjm,Hohm:2017pnh,Deser:2018oyg,Jurco:2019woz,Cederwall:2018aab,Cagnacci2019,Hohm:2019wql} for more recent reviews and applications).
Hence, it is tempting to construct an $L_\infty$-algebra from %in the infinity-enhanced Leibniz algebra extending 
a Leibniz algebra $V$. %The first steps of this direction were taken in 
Such a construction was presented in \cite{Kotov:2018vcz}, but 
%this construction produces an $L_\infty$-algebra that is freely generated, and thus 
it does not give back $V$ itself when $V$ is a Lie algebra.
%reduce to standard Yang-Mills gauge theory when $V$ is a mere Lie algebra. 
In \cite{Bonezzi:2019ygf}, a construction from an infinity-enhanced Leibniz algebra was given,
%strategy was used in 
but not
% is different: using the data contained in an infinity-enhanced Leibniz algebra, a construction of a $L_\infty$-algebra (that would lead to Yang-Mills theory when $V$ is a Lie algebra) is given}, but this construction was unfortunately not 
pushed further than to 
the 4-bracket.
In this paper however, using the fact that any Leibniz algebra canonically gives rise to a differential graded Lie algebra, we provide explicit formulas for the brackets of this $L_\infty$-algebra, to all orders. These formulas are obtained using a theorem by Getzler \cite{Getzler:2010}, which is a special case of a more general result obtained by Fiorenza and Manetti \cite{FiorenzaManetti}. This fact is another argument in favor of using differential graded Lie algebras over infinity-enhanced Leibniz algebras to encode tensor hierarchies.  The result also has the following consequence: that the skew-symmetric part of the Leibniz product of any Leibniz algebra $V$ can be lifted to an $L_\infty$-algebra structure in a canonical way, such that if the Leibniz product is fully skew-symmetric (\emph{i.e.}, if $V$ is a Lie algebra), then this \emph{$L_\infty$-extension} is $V$ itself.

The content of this paper can be summarized in the following diagram:
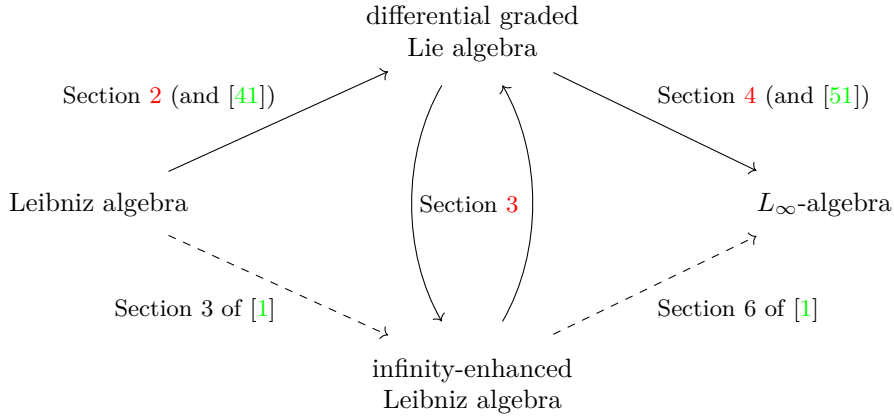
\vspace{3mm}
\begin{center}
%\begin{diagram}
\begin{tikzpicture}
\matrix(a)[matrix of math nodes, 
row sep=5em, column sep=6em, 
text height=1.5ex, text depth=0.25ex] 
{&\parbox{2.8cm}{\center differential graded Lie algebra}&\\
\text{Leibniz algebra}&&\text{$L_\infty$-algebra}\\ 
&\parbox{2.7cm}{\center infinity-enhanced Leibniz algebra}&\\}; 
%\path[right hook->](a-2-2) edge node[below right]{$\rho$} (a-1-3); 
\path[->, dashed, shorten >=0.6cm, shorten <=0.4cm](a-2-1) edge node[below left]{\footnotesize{Section 3 of \cite{Bonezzi:2019ygf}}} (a-3-2); 
\path[->, shorten >=0.6cm, shorten <=0.4cm](a-2-1) edge node[above left]{\footnotesize{Section \ref{embeddingleibniz}  (and \cite{Lavau:2017tvi})}} (a-1-2); %{\footnotesize{Sections \ref{triple} and \ref{sectionTHA}}} (a-1-2); 
\path[->, shorten >=0.5cm, shorten <=0.5cm](a-3-2) edge [bend right] node[left]{\footnotesize{Section \ref{infinityenhanced}}} (a-1-2); 
\path[->, shorten >=0.5cm, shorten <=0.5cm](a-1-2) edge [bend right] node{} (a-3-2); 
\path[->, dashed, shorten >=0.4cm, shorten <=0.6cm](a-3-2) edge node[below right]{\footnotesize{Section 6 of \cite{Bonezzi:2019ygf}}} (a-2-3); 
\path[->, shorten >=0.4cm, shorten <=0.6cm](a-1-2) edge node[above right]{\footnotesize{Section \ref{sectiongetzler} (and \cite{Getzler:2010})}} (a-2-3); 
%\path[->](a-2-1) edge node[above]{$\{\,.\,,.\,\}$} (a-2-2);
%\path[->](a-2-1) edge node[right]{$\Pi_W$} (a-1-1);
%\path[->](a-1-1) edge [bend right=60] node[left]{$\tau$} (a-3-1);
%\path[->](a-2-1) edge node[right]{} (a-3-1);
%\path[->](a-3-1) edge node[below right]{$\{\,.\,,.\,\}$} (a-2-2);
\end{tikzpicture}
%\end{diagram}
\end{center}
\vspace{3mm}
Dashed lines symbolize the results obtained in \cite{Bonezzi:2019ygf}, which were only partial: first, it was not shown that any Leibniz algebra induces an infinity-enhanced Leibniz algebra of order higher than 2 \cite{Strobl:2019hha}; second, the associated $L_\infty$-algebra was only given up to the 4-brackets.
On the other hand, the solid lines symbolize complete and canonical derivations that we will describe\footnote{The solid lines would correspond to functors in the language of category theory. Then, given the diagram, there would exist a canonical functor from the category of  Leibniz algebras to the category of $L_\infty$-algebras (that restricts to the identity functor on the full subcategory of Lie algebras).
%lifting the skew-symmetric part of the Leibniz bracket. 
However we do not address this question explicitly in this paper, because that would significantly increase its length and obscure our original motivation.  See also \cite{Kotov:2018vcz} for another point of view on this question.}. First, in Section \ref{embeddingleibniz}, we will explain the relation between Leibniz algebras and differential graded Lie algebras, and show that any Leibniz algebra $V$ canonically gives rise to a differential graded Lie algebra, which, in turn, induces a Leibniz product on $V$ that coincides with the original one. Second, in Section \ref{infinityenhanced} we explain in detail the 
%(not necessarily one-to-one) 
correspondence between differential graded Lie algebras and infinity-enhanced Leibniz algebras. It is functorial, but it does not define an equivalence of categories. Finally, in Section \ref{sectiongetzler}, we provide Getzler's theorem and the explicit formulas for the brackets of the $L_\infty$-algebra that was investigated in \cite{Bonezzi:2019ygf}. We end
Section \ref{sectiongetzler} with an application to the tensor hierarchy appearing in the $(1,0)$ superconformal model in six dimensions \cite{Samtleben:2011fj}.

To conclude, given that there is a correspondence between differential graded Lie algebras concentrated in non-negative degrees and infinity-enhanced Leibniz algebras, and although the latter notion arises more directly in the approach of \cite{Bonezzi:2019ygf}, we see several advantages to rather use the former %, rather than the latter, 
to mathematically encode tensor hierarchies:
%({\color{blue}although the latter arises more directly in the approach of \cite{Bonezzi:2019ygf}}):
\begin{itemize}
\item the algebraic structure of differential graded Lie algebras is much simpler;
\item any Leibniz algebra induces such a structure, in a canonical way;
\item such a structure induces an $L_\infty$-algebra, in a canonical and explicit way.
\end{itemize}
Moreover, the differential graded Lie algebra can in many interesting cases be seen as coming from a tensor hierarchy algebra \cite{Palmkvist:2013vya} (see footnote~1) which represents an
intriguing class of non-contragredient Lie superalgebras \cite{Carbone:2018njd}. Beyond the differential graded Lie algebra structure, the tensor hierarchy
algebras seem to
possess crucial information about supergravity and extended geometry
\cite{Greitz:2013pua,Bossard:2017wxl,Bossard:2019ksx,CederwallPalmkvistTHA}.

\section{Embedding tensors and Leibniz algebras}\label{embeddingleibniz}

A \emph{Leibniz algebra} is a  vector space $V$ equipped with a bilinear operation $\circ $ satisfying the derivation property, or \emph{Leibniz identity} \cite{Loday}:
\begin{equation}\label{leibnizidentity}
x\circ (y\circ  z)=(x\circ  y)\circ  z+y\circ (x\circ  z)
\end{equation}
for all $x,y,z\in V$.
We can  split the product $\circ$ of a Leibniz algebra $V$ into its symmetric part $\{.\,,.\}$ and its skew-symmetric part $[\,.\,,.\,]$:
\begin{equation}\label{splitting}
x\circ  y=\{x,y\}+[x,y]
\end{equation}
where
\begin{equation}
\{x,y\}=\frac{1}{2}\big(x\circ  y+y\circ  x\big) %\label{sym}
\hspace{0.7cm}\text{and}\hspace{0.7cm}
[x,y]=\frac{1}{2}\big(x\circ  y-y\circ  x\big)%\label{skewsym}
\end{equation}
for any $x,y\in V$.
As a consequence of \eqref{leibnizidentity}, the Leibniz product is a derivation of both brackets. 

The subspace $ U \subseteq V$ generated by the set of elements of the form $\{x,x\}$ contains all symmetric elements of the form $\{x,y\}$, since they can always be written as a sum of squares.
Using \eqref{leibnizidentity}, one can check that $ U $ is an ideal of $V$ with respect to the Leibniz product, {\it i.e.}, $V\circ  U \subseteq  U $.
We call this subspace the \emph{ideal of squares of $V$}. By \eqref{leibnizidentity}, the left action of $ U $ on $V$ is trivial:
\begin{equation}
 U \circ V=0.
\end{equation}
We say that this ideal is central, in the sense that it is included in the center $\mathcal{Z}$ of $V$ (with respect to the Leibniz product),
\begin{equation}
\mathcal{Z}=\Big\{x\in V\ \big|\ x\circ  y=0\ \ \text{for all}\ \ y\in V\Big\}.
\end{equation}

An important remark here is that even if the bracket $[\,.\,,.\,]$ is skew-symmetric, it does not satisfy the Jacobi identity since, using
\eqref{leibnizidentity}, we have
\begin{equation}\label{jacobiator0}
\big[x,[y,z]\big]+\big[y,[z,x]\big]+\big[z,[x,y]\big]=-\frac{1}{3}\Big(\big\{x,[y,z]\big\}+\big\{y,[z,x]\big\}+\big\{z,[x,y]\big\}\Big),
\end{equation}
which is not necessarily zero for all $x,y,z\in V$.  Hence the skew-symmetric bracket $[\,.\,,.\,]$ is not a Lie bracket, but since the left hand side of 
\eqref{jacobiator0} (the {\it Jacobiator}) takes values in the ideal of squares $U$, its action on $V$ is trivial.  

\subsection{From differential graded Lie algebras to Leibniz algebras}\label{sectiondgLa}

An important class of examples of Leibniz algebras come from {\it differential graded Lie algebras}, or \emph{dgLa} for short. A differential graded Lie algebra
\begin{align}
T=\cdots \oplus  T_{-2}\oplus T_{-1}\oplus T_0\oplus T_1\oplus T_2\oplus \cdots
\end{align}
 is in fact a $\mathbb{Z}$-graded Lie superalgebra
(and not a Lie algebra, as the term `differential graded Lie algebra' unfortunately suggests), where the $\mathbb{Z}$-grading is consistent with the $\mathbb{Z}_2$-grading. This means that it is equipped with a bilinear bracket which is graded skew-symmetric and satisfying the \emph{graded Jacobi identity}:
\begin{align}
\llbracket a,b \rrbracket &=-(-1)^{\ell(a)\ell(b)} \llbracket b,a \rrbracket,\label{gradedantisym}\\
\llbracket a,\llbracket b,c\rrbracket\rrbracket&=\llbracket \llbracket a,b\rrbracket,c\rrbracket+(-1)^{\ell(a)\ell(b)}\llbracket b,\llbracket a,c\rrbracket\rrbracket, \label{jacobi}
\end{align}
where $\ell(a)$ denotes the $\mathbb{Z}$-degree
of a homogeneous element $a \in T_{\ell(a)}$.
As a differential graded Lie algebra, $T$ is in addition equipped with a differential
\begin{align}
\partial=\big(\partial_{i}:T_{i+1}\to T_{i}\big){}_{i\in \mathbb{Z}},
\end{align}
which is an odd derivation of $T$ that squares to zero. The odd derivation property means that $\partial$ acts by the \emph{Leibniz rule}
\begin{align}\label{eq:derivationpartial}
\partial\big(\llbracket a,b\rrbracket\big)&=\llbracket\partial(a),b\rrbracket+(-1)^{\ell(a)}\llbracket a,\partial(b)\rrbracket.
\end{align}
One can then define the following degree $-1$ \emph{derived bracket} on $T_1$ \cite{Kosmann:2003}:
\begin{align}\label{eq:derivedbracket}
x\circ y\equiv \llbracket\partial(x),y\rrbracket
\end{align}
for any $x,y\in T_1$. The Jacobi identity (\ref{jacobi}) and the Leibniz rule (\ref{eq:derivationpartial}) together ensure that this product is a derivation of itself, \emph{i.e.}, that it is a Leibniz product on $T_1$.

In the category of graded vector spaces, the grading of $T$ can be shifted by $-1$ by using the \emph{suspension operator $s$}.
It is defined as follows:\footnote{Obviously, the suspension operator has an inverse. It is called the \emph{desuspension operator} and is denoted $s^{-1}$.}
\begin{align}
(sT)_i\equiv T_{i+1}
\end{align}
In the present paper, since using too much mathematical notations could be obfuscating, we will not write $s(a)$ for the suspension of an element $a\in T$, but we will stick to the notation~$a$.
However, to avoid any confusion between the grading on  $T$ and the grading on $sT$, we choose to denote by $\ell(a)$ the degree of $a$ as seen as an element of $T$, and $|a|$ the degree of $a$ as seen as an element of $sT$. Hence:
\begin{align}
|a|=\ell(a)-1
\end{align}

From now on, we assume (if nothing else explicitly stated)
that the differential graded Lie algebras are concentrated in non-negative degrees, \emph{i.e.},
$T=T_0\oplus T_1\oplus T_2\oplus \cdots$. We also define $\overline{T}$ to be the subspace of $T$ that consists of strictly positively graded elements, \emph{i.e.},
\begin{align}
\overline{T}=T_1\oplus T_2\oplus\cdots
\end{align}
We call $\textbf{dgLa}_{\geq0}$ the category of non-negatively graded differential graded Lie algebras, with their associated morphisms.  
Then, we may identify the differential $\partial$ with the adjoint action of an element, which we denote by $\Theta$, in an additional one-dimensional subspace
$T_{-1}$ that we may add to $T$ by a direct sum, such that
\begin{align}\label{eq:quadratic}
\llbracket \Theta,\Theta\rrbracket=0.
\end{align}
This quadratic constraint on $\Theta$ ensures that the operator $\llbracket\Theta,-\rrbracket: T_\bullet\to T_{\bullet -1}$ squares to zero. The derivation property \eqref{eq:derivationpartial} translates to a Jacobi identity that $\Theta$ has to satisfy:
\begin{align}\label{eq:jacobitheta}
\llbracket \Theta,\llbracket a,b\rrbracket\rrbracket&=\llbracket \llbracket \Theta,a\rrbracket,b\rrbracket+(-1)^{\ell(a)}\llbracket a,\llbracket \Theta,b\rrbracket\rrbracket
\end{align}
for any $a,b\in T$. Hence the differential graded Lie algebra structure on $T_0\oplus T_1\oplus\cdots$ can be interpreted as a 
$\mathbb{Z}$-graded Lie superalgebra structure on $T_{-1}\oplus T_0\oplus T_1\oplus\cdots$.
With this identification, (\ref{eq:derivedbracket}) implies
\begin{align}\label{eq:moduleV}
\llbracket \llbracket \Theta,x\rrbracket,y\rrbracket = x\circ y.
\end{align}

The reverse construction, \emph{i.e.}, starting from a Leibniz algebra and building a
differential graded Lie algebra such that the induced Leibniz product \eqref{eq:moduleV} coincides with the original one, has been given in \cite{Lavau:2017tvi}. It relies on the notion of \emph{Lie-Leibniz triples} and has been inspired by the gauging procedure in supergravity, in particular by the embedding tensor formalism
and the correspondence between embedding tensors and Leibniz algebras \cite{Hohm:2018ybo,Kotov:2018vcz}.
This will be the topic of the next two subsections.

\subsection{Embedding tensors and Lie-Leibniz triples}\label{triple}

An example of a Leibniz algebra arises in the embedding tensor approach to gauged supergravity \cite{deWit:2002vt,deWit:2004nw,
deWit:2005hv,deWit:2008ta,Trigiante:2016mnt}.
In the gauging procedure, a subgroup $H$ of the global duality symmetry group $G$
is promoted to a local symmetry group.
Covariance under $G$ can be maintained with the help of an \emph{embedding tensor}, which is a linear map $\Theta:V\to \mathfrak{g}$ from a $\mathfrak{g}$-module $V$ (usually fundamental) to the Lie algebra $\mathfrak{g}$ of the original global symmetry group $G$, describing how $H$ is embedded into $G$. For consistency, this embedding tensor has to satisfy a \emph{representation constraint} and a \emph{quadratic constraint}. One may now define a Leibniz algebra structure on $V$ by using the action of $\mathfrak{h}=\mathrm{Im}(\Theta)\subseteq \mathfrak{g}$ on $V$:
\begin{align}\label{condition3}
x\circ y\equiv\rho_{\Theta(x)}(y),
\end{align}
where the representation
$\rho:\mathfrak{g}\to \mathfrak{gl}(V)$ defines the $\mathfrak{g}$-module structure on $V$. The Leibniz identity is then a consequence of the quadratic constraint
\begin{align}\label{condition4}
\Theta\big(\rho_{\Theta(x)}(y)\big)=\big[\Theta(x),\Theta(y)\big]_{\mathfrak{g}}
\end{align}
since
\begin{align}
x \circ (y \circ z) -  y \circ (x \circ z) &= \rho_{\Theta(x)}\big(\rho_{\Theta(y)}(z)\big)-\rho_{\Theta(y)}\big(\rho_{\Theta(x)}(z)\big)\nonumber\\
  &=\big[\rho_{\Theta(x)},\rho_{\Theta(y)}\big]_{\mathfrak{gl}(V)}(z)\nonumber\\
  &=\rho_{[\Theta(x),\Theta(y)]_\mathfrak{g}}(z)\nonumber\\
  &=\rho_{\Theta(\rho_{\Theta(x)}(y))}(z)\nonumber\\
  &=\rho_{\Theta(x \circ y)}(z)=(x\circ y) \circ z.
\end{align}

Conversely, one can show that a Leibniz algebra $(V,\circ)$ canonically defines an embedding tensor, since any vector space $V$ is a $\mathfrak{g}$-module
with
$\mathfrak{g}\equiv \mathfrak{gl}(V)$
and $\rho$ being the identity map.
We then let $\Theta$ be the map sending any element $x\in V$ to its associated left-multiplication operator $x_L$:
\begin{align}
\Theta:\hspace{0.2cm}V&\xrightarrow{\hspace*{1.2cm}} \hspace{0.2cm}\mathfrak{gl}(V)\label{eq:embeddingtensor}\\
	x&\xmapsto{\hspace*{1.2cm}}x_L:y\mapsto x\circ y.\nonumber
\end{align}
The image of the map $\Theta$ is a subspace $\mathfrak{h}$ of $\mathfrak{gl}(V)$ generated by all endomorphisms of the type $x_L$ for $x\in V$. 
It turns out to be stable under the Lie bracket of endomorphisms $[\,.\,,.\,]_{\mathfrak{gl}(V)}$, since the Leibniz identity \eqref{leibnizidentity} implies that $[x_L,y_L]_{\mathfrak{gl}(V)}=(x\circ y)_L$.
This equality can be rewritten as the condition that $\Theta:V\to \mathfrak{gl}(V)$ is a homomorphism of Leibniz algebras:
\begin{align}\label{condition1}
\Theta(x\circ y)=\big[\Theta(x),\Theta(y)\big]_{\mathfrak{gl}(V)}
\end{align}
where one considers $\big(\mathfrak{gl}(V),[\,.\,,.\,]_{\mathfrak{gl}(V)}\big)$ as a Leibniz algebra with fully skew-symmetric product.
One can thus restrict the action of $\mathfrak{gl}(V)$ on $V$ to $\mathfrak{h}$, turning $V$ into a $\mathfrak{h}$-module. The Leibniz product is then compatible with the embedding tensor in the same sense as in \eqref{condition3}:
\begin{align}\label{condition2}
\Theta(x)(y)=x\circ y.
\end{align}
Inserting \eqref{condition2} into \eqref{condition1} implies that $\Theta$ satisfies the quadratic constraint \eqref{condition4}.
Moreover, from \eqref{condition1} we see that the kernel of $\Theta$ coincides with the center $\mathcal{Z}$ of $V$. This means that the gauge algebra $\mathfrak{h}=\mathrm{Im}(\Theta)$ is isomorphic to the Lie algebra $\bigslant{V}{\mathcal{Z}}$.

We have seen that any embedding tensor defines a Leibniz algebra, and that any Leibniz algebra defines an embedding tensor in a canonical way. 
One can generalize the discussion and allow for more general Lie algebras and embedding tensors, that would satisfy the two constraints \eqref{condition1} and \eqref{condition2}. These are consistency conditions that encode compatibility between the embedding tensor and the Leibniz algebra structure \cite{Lavau:2017tvi}. 

\begin{definition*}
A \emph{Lie-Leibniz triple} is a triple $(\mathfrak{g},V,\Theta)$ where:
\begin{enumerate} 
\item $\mathfrak{g}$ is a Lie algebra,
\item $V$ is a $\mathfrak{g}$-module equipped with a Leibniz algebra structure $\circ$, and
\item $\Theta: V\to \mathfrak{g}$ is a linear map called the \emph{embedding tensor}, that satisfies two compatibility conditions.  The first one is the \emph{linear constraint}:
\begin{equation}\label{eq:compat}
x\circ y=\rho_{\Theta(x)}(y)
\end{equation}
where $\rho:\mathfrak{g}\to \mathfrak{gl}(V)$ denotes the action of $\mathfrak{g}$ on $V$.
The second one is called the \emph{quadratic constraint}:
\begin{equation}\label{eq:equiv}
\Theta(x\circ  y)=\big[\Theta(x),\Theta(y)\big]_{\mathfrak{g}}
\end{equation}
where $[\,.\,,.\,]_{\mathfrak{g}}$ is the Lie bracket on $\mathfrak{g}$.
\end{enumerate}
\end{definition*}

The two conditions that $\Theta$ has to satisfy guarantee the compatibility between the Leibniz algebra, $\mathfrak{g}$-module structure on $V$
and the Lie bracket of $\mathfrak{g}$. 
In particular, the quadratic constraint (\ref{eq:equiv}) says that $\Theta$
has to be a homomorphism of Leibniz algebras, considering the Lie algebra $\mathfrak{g}$ as a Leibniz algebra whose product is fully skew-symmetric. Given these data, we deduce that $\mathfrak{h}\equiv\mathrm{Im}(\Theta)$ is a Lie subalgebra of $\mathfrak{g}$. We call it the \emph{gauge algebra of the Lie-Leibniz triple $(\mathfrak{g},V,\Theta)$}. 
Moreover, the linear constraint \eqref{eq:compat} implies that the kernel of $\Theta$ is contained in the center,
\begin{equation}\label{inclusion}
\mathrm{Ker}(\Theta)\subseteq \mathcal{Z}.
\end{equation}
We have equality when the representation of $\mathfrak{h}$ on $V$ is faithful (as in the case of the embedding tensor canonically defined by any Leibniz algebra as in \eqref{eq:embeddingtensor}). On the other hand,
the quadratic constraint \eqref{eq:equiv} implies that the
ideal of squares $U$ is contained in the kernel,
\begin{equation}
 U \subseteq\mathrm{Ker}(\Theta),
\end{equation}
but {\it a priori} we do not have equality here either.

\subsection{From Leibniz algebras to differential graded Lie algebras}\label{sectionTHA}

In \cite{Lavau:2017tvi} it was shown that a Lie-Leibniz triple $(\mathfrak{g},V,\Theta)$ gives rises to a 
dgLa
\begin{align}
T = T_0 \oplus T_1 \oplus T_2 \oplus \cdots,
\end{align}
which satisfies several properties that makes it unique (up to equivalence).
Among other properties \cite{Lavau:2017tvi}, the differential graded Lie algebra $\big(T,\llbracket\,.\,,.\,\rrbracket,\partial\big)$ induced by the Lie-Leibniz triple $(\mathfrak{g},V,\Theta)$ 
is such that 
\begin{enumerate}
\item the subalgebra $T_0$ is equal to $\mathfrak{h}$ as a Lie algebra,
\item the space $T_1$ is equal to $V$,
\item for any $i\geq1$, the space $T_i$ is a $\mathfrak{g}$-module with a representation $\rho$ given by
\begin{align}
\llbracket g,a\rrbracket\equiv \rho_g (a), \label{T1=V}
\end{align}
for any $g \in \mathfrak{h}=T_0$ and $a \in T_i$,
\item the differential satisfies
\begin{align}
\partial(x)\equiv\Theta(x) \label{partial=theta}
\end{align}
for any $x\in T_1$.
\end{enumerate}
The combination of (\ref{T1=V}) and (\ref{partial=theta}), together with the fact that the embedding tensor $\Theta$ satisfies \eqref{eq:compat}, imply the following identity:
\begin{equation}\label{adjointaction2}
\llbracket \partial (x),y\rrbracket=\llbracket \Theta(x),y\rrbracket=\rho_{\Theta(x)}(y)=x\circ y
\end{equation}
for any $x,y\in T_1=V$.  Hence, the Leibniz product defined by the dgLa structure as in \eqref{eq:moduleV} coincides with the original one, and this is a particular feature of this differential graded Lie algebra.
Since any Leibniz algebra $V$ canonically induces a Lie-Leibniz triple $(\mathfrak{gl}(V),V,\Theta:x\mapsto x_L)$, we deduce that any Leibniz algebra gives rise to such a dgLa.

Schematically, the construction in \cite{Lavau:2017tvi}
of the dgLa $T$ associated to
the Lie-Leibniz triple $(\mathfrak{g},V,\Theta)$ goes as follows: one sets $X_0=V$, then one chooses $X_1$ by noticing that the symmetric bracket $\{\,.\,,.\,\}:S^2(V)\to V$ has a kernel, which is a $\mathfrak{h}$-module (where $\mathfrak{h}=\mathrm{Im}(\Theta)$), but not necessarily a $\mathfrak{g}$-module. Let $K\subseteq\mathrm{Ker}\big(\{\,.\,,.\,\}\big)$ be the biggest $\mathfrak{g}$-module contained in $\mathrm{Ker}\big(\{\,.\,,.\,\}\big)$. Then one sets $X_1={S^2(V)}\big/{K}$ and $X_1$ naturally inherits the quotient $\mathfrak{g}$-module structure. Usually, $S^2(V)$ is completely reducible with respect to the action of $\mathfrak{g}$ so that the quotient is a direct sum of irreducible representations. In that case, $X_1$ is the smallest $\mathfrak{g}$-submodule of $S^2(V)$ through which the symmetric bracket $\{\,.\,,.\,\}:S^2(V)\to V$ factors:
\vspace{2mm}
\begin{center}
\begin{tikzpicture}
\matrix(a)[matrix of math nodes, 
row sep=5em, column sep=6em, 
text height=1.5ex, text depth=0.25ex] 
{&X_1\\ 
S^2(V)&V\\}; 
\path[->>](a-2-1) edge node[above left]{$\mathbb{P}_1$} (a-1-2);  
\path[->](a-1-2) edge node[right]{} (a-2-2);
\path[->](a-2-1) edge node[above]{$\{.\,,.\}$} (a-2-2);
\end{tikzpicture}
\vspace{2mm}
\end{center}

\bigskip
Here $\mathbb{P}_1$ is the projection on $X_1$, and the vertical arrow is uniquely defined by requiring that the diagram commutes, given the projection $\mathbb{P}_1$.  Elements of $X_1$ are considered to have degree $+1$. In gauged supergravity, $X_1$ is the space in which 2-form potentials take values and is determined by the representation constraint. More generally, in the hierarchy, $X_p$ is the $\mathfrak{g}$-module in which $(p+1)$-form potentials take values. 
We extend $\mathbb{P}_1$ to a map $\widetilde{\mathbb{P}}_1:S^\bullet(V)\to S^{\bullet-2}(V)\otimes X_1$ and we set $X_2$ to be the cokernel of this map when applied to~$S^3(V)$:

\begin{center}
\begin{tikzpicture}
\matrix(a)[matrix of math nodes, 
row sep=5em, column sep=6em, 
text height=1.5ex, text depth=0.25ex] 
{&&X_2\\
&V\otimes X_1&X_1\\ 
S^3(V)&S^2(V)&V\\}; 
\path[->>](a-3-2) edge node[above left]{$\mathbb{P}_1$} (a-2-3);
\path[->>](a-2-2) edge node[above left]{$\mathbb{P}_2$} (a-1-3);
\path[->](a-3-1) edge node[above left]{$\widetilde{\mathbb{P}}_1$} (a-2-2);
\path[->](a-1-3) edge node[right]{} (a-2-3);
\path[->](a-2-3) edge node[right]{} (a-3-3);
\path[->](a-3-2) edge node[above]{$\{.\,,.\}$} (a-3-3);
\end{tikzpicture}
\end{center}

\bigskip
Here $\mathbb{P}_2$ is the quotient map onto $X_2$. 
We extend it to $\widetilde{\mathbb{P}}_2 :S(V\oplus X_1)\to S(V\oplus X_1)\otimes X_2$ so that we can define $X_3$ as the cokernel of $\widetilde{\mathbb{P}}_1+\widetilde{\mathbb{P}}_2:S^2(V)\otimes X_1\to V\otimes X_2\oplus X_1\vee X_1$. We repeat the same construction at each iteration, and we obtain a tower of graded spaces $X=X_0\oplus X_1\oplus X_2\oplus \cdots$, together with their respective projections: $\mathbb{P}_i:S^2(X)\to X_i$. The detailed construction of the hierarchy can be found in \cite{Lavau:2017tvi}. 
Notice that if $V$ is a  Lie algebra, then the kernel of the symmetric bracket coincides with $S^2(V)$ and then $X_1$ is the zero vector space. This implies in turn that all other spaces $X_p$ are zero. 

The differential graded Lie algebra $T$ is obtained as follows: one first shifts the degree of each space by $+1$, \emph{i.e.}, $\overline{T}=s^{-1}X$. In particular we have $T_1=V$ and $T_i=X_{i-1}$ for any $i\geq1$. Then, we add $\mathfrak{h}=\mathrm{Im}(\Theta)$ at degree $0$, \emph{i.e.}, $T_0=\mathfrak{h}$, with its associated Lie bracket. After some technical considerations, the projection maps $\mathbb{P}_i$ induce a graded Lie bracket $\llbracket\,.\,,.\,\rrbracket$ on $T=T_0\oplus T_1\oplus T_2\oplus \cdots$. One can show that the vertical arrows in the above diagram  canonically define a differential $\partial$ on $T$ such that the triple $\big(T,\llbracket\,.\,,.\,\rrbracket,\partial\big)$ is a differential graded Lie algebra \cite{Lavau:2017tvi}. Notice that the presence of $T_0=\mathfrak{h}$ is crucial so that the Leibniz rule \eqref{eq:derivationpartial} is satisfied. If $V$ is a Lie algebra, then $T=T_0\oplus T_1$ and there is no space of degree higher than or equal to 2.

\subsection{Construction from a universal $\mathbb{Z}$-graded Lie superalgebra}

Another way of constructing a dgLa associated to a Leibniz algebra $V$ (corresponding to the canonical Leibniz-Lie triple) is to first consider the universal
$\mathbb{Z}$-graded Lie superalgebra $\mathcal{U}(V)$
associated to $V$, where $V$ is considered as a $\mathbb{Z}_2$-graded vector space with a trivial even part
\cite{Kantor-graded,Palmkvist:2009qq,Palmkvist:2013vya}. This means that we set $\mathcal{U}_1=V$
and define vector spaces $\mathcal{U}_{-i}$ for $i\geq0$ recursively so that $\mathcal{U}_{-i}=\mathrm{Hom}(V,\mathcal{U}_{-i+1})$,
consisting of all linear maps $V\to\mathcal{U}_{-i+1}$.
Then the direct sum
$\bigoplus_{i\geq0}\mathcal{U}_{-i}$
is a consistently $\mathbb{Z}$-graded Lie superalgebra with the Lie superbracket defined recursively by
\begin{align}
\llbracket a,b \rrbracket (x)= \llbracket a, b(x)\rrbracket - (-1)^{\ell{(a)}\ell{(b)}} \llbracket b, a(x)\rrbracket,
\end{align}
where
$\ell(a)$ is the $\mathbb{Z}$-degree of a homogeneous element $a \in \mathcal{U}_{\ell(a)}$, and $\llbracket a , x \rrbracket$ should be read as $a(x)$
if $x\in\mathcal{U}_1=V$ and $\ell(a) \leq 0$.
In particular, this means that the subalgebra $\mathcal{U}_0$ is $\mathfrak{gl}(V)$.
The Jacobi identity can then be shown to hold by induction. 
We can extend this Lie superalgebra to positive degrees as well, by letting
$\bigoplus_{i\geq1}\mathcal{U}_{i}$ be the free Lie superalgebra generated by the odd vector space $\mathcal{U}_1=V$.
The Lie superbracket of an element at a positive degree with an element at a non-positive degree can be defined by the Jacobi identity from the relations
\begin{align}
\llbracket a , x \rrbracket = -\llbracket x , a \rrbracket = a(x)
\end{align}
for $x\in\mathcal{U}_1=V$ and $\ell(a) \leq 0$.
In this way we obtain a consistently
$\mathbb{Z}$-graded Lie superalgebra
\begin{align}
\mathcal{U}(V) = \bigoplus_{i\geq0}\mathcal{U}_{-i} \oplus \bigoplus_{i\geq1}\mathcal{U}_{i} = \bigoplus_{i\in \mathbb{Z}}\mathcal{U}_{i}
\end{align}
for any vector space $V$. When $V$ happens to be an algebra with a product $x \circ y$, there is a distinguished element $\Theta$ in $\mathcal{U}_{-1}$ defined
by $\llbracket \llbracket \Theta , x \rrbracket,y \rrbracket= x \circ y$ for any $x,y \in \mathcal{U}_1=V$, and the condition that
$V$ be a Leibniz algebra is then equivalent to the condition $\llbracket \Theta,\Theta \rrbracket =0$.

Consider now the subalgebra $R=\bigoplus_{i\in \mathbb{Z}}R_i$ of $\mathcal{U}(V)$ generated by $V=\mathcal{U}_1$
and $\Theta \in \mathcal U_{-1}$. At degree $0$ in $R$, we have all linear maps of the form $\llbracket \Theta, x \rrbracket=x_L$
for $x \in V$, acting on $y\in V$ by $x_L(y)=x \circ y$. The bracket of any such element $x_L$ with $\Theta$ does not give any new elements at degree $-1$
since
\begin{align}
\llbracket \Theta, x_L \rrbracket = \llbracket \Theta, \llbracket \Theta, x \rrbracket\rrbracket = \frac12 \llbracket\llbracket \Theta,\Theta \rrbracket,x \rrbracket
=0.
\end{align} 
Thus $R_{-1}$ is one-dimensional, spanned by $\Theta$, and $R_{-i}$ for $i\geq2$ are trivial since $\llbracket \Theta,\Theta \rrbracket=0$.
Thus the subalgebra $R$ of $\mathcal{U}(V)$ generated by $V=\mathcal{U}_1$
and $\Theta \in \mathcal U_{-1}$ is a Lie superalgebra concentrated in degrees $\geq -1$ with a one-dimensional subspace $\mathcal{U}_{-1}$.
As we have seen in Section \ref{sectiondgLa}, it can then be identified with a dgLa concentrated in non-negative degrees.
The dgLa constructed from the canonical Lie-Leibniz triple in Section \ref{sectionTHA} may then be obtained as a quotient.
We leave the study of the precise relation for future work.

The tensor hierarchy algebra introduced in \cite{Palmkvist:2013vya} is defined in a similar way, from the universal $\mathbb{Z}$-graded Lie superalgebra associated to
a $\mathfrak{g}$-module $V$, but in that construction the relevant subalgebra is not generated by $V=\mathcal{U}_1$ and a single element $\Theta \in \mathcal{U}_{-1}$,
but by $V=\mathcal{U}_1$ and a whole subspace of $\mathscr{U}_{-1}$, which is the $\mathfrak{g}$-module containing all allowed
embedding tensors according to the
representation constraint. The tensor hierarchy algebra is then obtained by factoring out the maximal ideal of this subalgebra contained in the subspaces at degree $2$ and higher. Choosing a particular embedding tensor $\Theta \in\mathcal{U}_{-1}$ amounts to defining a Leibniz algebra structure on $V$ and restricting the tensor hierarchy algebra to a dgLa. This dgLa then coincides with the one constructed from the Lie-Leibniz triple $(\mathfrak{g},V,\Theta)$ in Section \ref{sectionTHA} up to possible differential ideals in the latter contained in the subspaces at degree strictly higher than 2.

\section{Infinity-enhanced Leibniz algebras}\label{infinityenhanced}

In \cite{Strobl:2016aph,Strobl:2019hha},
the concept of an {\it enhanced Leibniz algebra} was introduced as a first step towards a mathematical formalization of higher gauge theories. It did not allow 
for gauge fields of form degree higher than 2 though, so that 
the notion of an {\it infinity-enhanced Leibniz algebra} was eventually proposed in  \cite{Bonezzi:2019ygf} as the most general structure encoding
the tensor hierarchy.

\begin{definition*}
An \emph{infinity-enhanced Leibniz algebra} is defined as an $\mathbb{N}$-graded vector space
\begin{align}
X=X_0 \oplus X_1 \oplus X_2 \oplus \cdots = X_0 \oplus \overline X
\end{align}
together with a Leibniz product $\circ : X_0 \otimes X_0 \to X_0$,
\begin{equation}\label{leibnizidentity2}
x\circ (y\circ  z)=(x\circ  y)\circ  z+y\circ (x\circ  z),
\end{equation}
a degree $+1$ graded symmetric product $\bullet:X_{i}\otimes X_j \to X_{i+j+1}$ ($i,j\geq 0$),
\begin{align} \label{eq:definitionbullet}
a\bullet b=(-1)^{|a||b|}b\bullet a
\end{align}
(where $|a|$ denotes the $\mathbb{Z}$-degree of a homogeneous element $a \in X_{|a|}$) and
a linear map $\mathfrak{D}=\big(\mathfrak{D}_{i}:X_{i+1}\to X_{i}\big){}_{i\geq0}$
satisfying in addition the following axioms:
\begin{enumerate}
\item $\mathfrak{D}u \circ x =0 \qquad\! (u \in X_1,\,x \in X_0)$,
\item $\mathfrak{D}(x\bullet y)=x\circ y+y\circ x \qquad (x,y \in X_0)$,
\item $\mathfrak{D}\big(x\bullet (y\bullet z)\big)=(x\circ y)\bullet z+(x\circ z)\bullet y-(y\circ z+z\circ y)\bullet x \qquad (x,y,z\in X_0)$,
\item $\mathfrak{D}\big(x_{[1}\bullet (x_{2]}\bullet u)\big)=2\, x_{[2}\bullet\mathfrak{D}\big(x_{1]}\bullet u\big)+x_{[2}\bullet \big(x_{1]}\bullet \mathfrak{D}(u)\big)+[x_1,x_2]\bullet u\quad  (x_1,x_2\in X_0,\, u\in\overline{X})$,
\item
$\mathfrak{D}(u\bullet v)+(\mathfrak{D}u)\bullet v+(-1)^{|u|}u\bullet \mathfrak{D}(v)=0 \qquad (u,v\in \overline{X})$,
\item 
$(-1)^{|a|}a\bullet(b\bullet c)+ (a\bullet b)\bullet c + (-1)^{|b||c|}(a\bullet c)\bullet b=0 \qquad (a,b,c \in X)$,
\item $\mathfrak{D}\big(\mathfrak{D}(a))=0\qquad (a\in X)$.
\end{enumerate}
where in the latter condition, $\mathfrak{D}$ is assumed to vanish on $X_0$. 

A {\rm morphism of infinity-enhanced Leibniz algebras} is a degree preserving linear map that is compatible in the usual way with the products and the differentials. More precisely, for two infinity-enhanced Leibniz algebra $(X, \circ, \mathfrak{D}, \bullet)$ and $(Y, \circ', \mathfrak{D}', \bullet')$, a morphism  $f:X\to Y$ satisfies the following three conditions $(a,b\in X)$:
\begin{equation}
f(a\circ b)=f(a)\circ' f(b), \qquad f(a\bullet b) = f(a) \bullet' f(b), \qquad f\big(\mathfrak{D}(a)\big)=\mathfrak{D}'\big(f(a)\big).\label{inftymorphism}
\end{equation}

Let $\mathrm{\infty\mathbf{-enLeib}}$ be the category of infinity-enhanced Leibniz algebras, with the above defined morphisms.
\end{definition*}

An important operator introduced in \cite{Bonezzi:2019ygf} is the generalized Lie derivative of an element $x$ of $X_0$. It acts on the entirety of the chain complex $X=X_0\oplus \overline{X}$ and is defined by the following two equations:
\begin{align}
\mathcal{L}_x(y)&\equiv x\circ y \qquad (y\in X_0),\label{Leibnizhigher2}\\
\mathcal{L}_x(u)&\equiv x\bullet \mathfrak{D}(u)+\mathfrak{D}(x\bullet u)\qquad (u\in \overline{X}).\label{Leibnizhigher}
\end{align}
It defines an action of $X_0=V$ on each subspace $X_i$, and it turns out that the above axioms imply that any previously defined operators $\circ,\mathfrak{D},\bullet$ are covariant under the action of this generalized Lie derivative. That is to say,
\begin{align}
\mathcal{L}_x(y\circ z)&=\mathcal{L}_x(y)\circ z+y\circ \mathcal{L}_x(z)\qquad(y,z\in X_0),\\
\mathcal{L}_x\big(\mathfrak{D}(u)\big)&=\mathfrak{D}\big(\mathcal{L}_x(u)\big)\qquad(u\in\overline{X}),\\
\mathcal{L}_x(a\bullet b)&=\mathcal{L}_x(a)\bullet b+a\bullet \mathcal{L}_x(b)\qquad(a,b\in X).
\end{align}
Moreover, this generalized Lie derivative automatically satisfies a \emph{closure condition}:
\begin{align}\label{closurecondition}
\big[\mathcal{L}_x,\mathcal{L}_y\big]=\mathcal{L}_{[x,y]}
\end{align}
where, on the right hand side, $[x,y]$ is the skew-symmetric part of the Leibniz product $x\circ y$
(and the left hand side is just a commutator of linear maps).

This section is devoted to showing that any dgLa $T=T_0\oplus T_1\oplus T_2\oplus\cdots=T_0\oplus \overline{T}$ canonically induces an infinity-enhanced Leibniz algebra structure on $X=s\,\overline{T}$
and, conversely,
that any infinity-enhanced Leibniz algebra $X=X_0\oplus X_1\oplus X_2\oplus \cdots$ canonically gives rise to a dgLa structure on $T=T_0\oplus s^{-1}X$, where $T_0$ is the image of the embedding tensor defined in \eqref{eq:embeddingtensor}.

First, let us show that the axioms above follow from any dgLa $T=T_0\oplus T_1\oplus T_2\oplus\cdots$
(extended to a graded Lie algebra $T_{-1} \oplus T_0 \oplus T_1 \oplus \cdots$ with a one-dimensional subspace $T_{-1}$, identifying the differential with
a basis element $\Theta$ of $T_{-1}$ squaring to zero, see \eqref{eq:quadratic}).
To this end, for every $i\geq0$, we set $X_i=T_{i+1}$, as well as the following operations:
\begin{align}
x \circ y &\equiv \llbracket \llbracket \Theta, x \rrbracket , y \rrbracket \qquad (x,y \in T_1=s^{-1}X_0),
\label{eq:defcircle}\\
a\bullet b&\equiv (-1)^{\ell(a)}\llbracket a,b\rrbracket\qquad (a,b \in \overline T=s^{-1}X),
\label{eq:defbullet}\\
\mathfrak{D}&\equiv-\llbracket\Theta,\,-\,\rrbracket
\label{eq:defdiff}
\end{align}
where elements on the left hand side are considered as elements of $X$, whereas elements on the right hand side are considered as elements of $\overline{T}$. 
It then follows that (\ref{eq:definitionbullet}) is satisfied:
\begin{align}
a\bullet b &=(-1)^{\ell(a)}\llbracket a,b\rrbracket\nonumber\\
&=-(-1)^{\ell(b)\ell(a)+\ell(a)}\llbracket b,a\rrbracket\nonumber\\
&=(-1)^{(\ell(a)-1)(\ell(b)-1)+\ell(b)}\llbracket b,a\rrbracket\nonumber\\
&=(-1)^{|a||b|} b\bullet a
\end{align}
since $|a|=\ell(a)-1$ (and likewise for $b$). The identity \eqref{leibnizidentity2} follows from the Leibniz rule \eqref{eq:derivationpartial} and the Jacobi identity \eqref{eq:jacobitheta}, and it is easy to see that the other basic conditions on $\circ$, $\bullet$ and $\mathfrak{D}$ are satisfied. Let us now go through the additional six axioms. 

\textbf{Axiom 1:} By \eqref{eq:defcircle} and \eqref{eq:defdiff} we have
 \begin{align}
 \mathfrak{D}u\circ x=\llbracket\llbracket\Theta,\mathfrak{D}u\rrbracket,x\rrbracket=-\llbracket\llbracket\Theta,\llbracket\Theta,u\rrbracket\rrbracket,x\rrbracket=-\frac{1}{2}\llbracket\llbracket\llbracket\Theta,\Theta\rrbracket,u\rrbracket,x\rrbracket,
 \end{align}
 which vanishes by \eqref{eq:quadratic}.

 \textbf{Axiom 2:} Using \eqref{eq:defbullet} and \eqref{eq:defdiff}, and recalling that $\ell(x)=1$, the left hand side of Axiom~2 is $\llbracket\Theta,\llbracket x, y\rrbracket\rrbracket$. By the Jacobi identity \eqref{eq:jacobitheta} on $T$, we have
 \begin{align}
 \llbracket\Theta,\llbracket x, y\rrbracket\rrbracket=\llbracket\llbracket\Theta, x\rrbracket, y\rrbracket
-\llbracket x,\llbracket\Theta, y\rrbracket\rrbracket=\llbracket\llbracket\Theta, x\rrbracket, y\rrbracket
+\llbracket \llbracket\Theta, y\rrbracket,x\rrbracket.
 \end{align}
 Using \eqref{eq:defcircle}, this is equal to $x\circ y+y\circ x$, which is precisely the right hand side of Axiom 2.

\textbf{Axiom 3:} Using \eqref{eq:defbullet} and \eqref{eq:defdiff}, the left hand side of Axiom 3 is $-\llbracket\Theta,\llbracket x,\llbracket y,z\rrbracket\rrbracket\rrbracket$. Then the Jacobi identity \eqref{eq:jacobitheta} implies
\begin{align}
-\llbracket\Theta,\llbracket x,\llbracket y,z\rrbracket\rrbracket\rrbracket&=-\llbracket\llbracket\Theta,x\rrbracket,\llbracket y,z\rrbracket\rrbracket+\llbracket x,\llbracket \llbracket\Theta,y\rrbracket,z\rrbracket\rrbracket-\llbracket x,\llbracket y,\llbracket\Theta,z\rrbracket\rrbracket\rrbracket\nonumber\\
&=-\llbracket\llbracket\llbracket\Theta,x\rrbracket,y\rrbracket,z\rrbracket-\llbracket y,\llbracket \llbracket\Theta,x\rrbracket,z\rrbracket\rrbracket\nonumber\\
&\hspace{3cm}+\llbracket \llbracket \llbracket\Theta,y\rrbracket,z\rrbracket,x\rrbracket+\llbracket \llbracket \llbracket\Theta,z\rrbracket,y\rrbracket,x\rrbracket\nonumber\\
&=-\llbracket x\circ y,z\rrbracket-\llbracket y,x\circ z\rrbracket+\llbracket y\circ z,x\rrbracket+\llbracket z\circ y,x\rrbracket,\label{equationaxiom3}
\end{align}
where we used \eqref{eq:defcircle} between the second line and the last one. By using \eqref{eq:defbullet}, the right hand side of \eqref{equationaxiom3} can be written as $(x\circ y)\bullet z+ (x\circ z)\bullet y-(y\circ z+z\circ y)\bullet x$, which is the right hand side of Axiom 3.

\textbf{Axiom 4:}  Using \eqref{eq:defbullet} and \eqref{eq:defdiff}, the left hand side of Axiom 4 is $-\llbracket\Theta,\llbracket x_{[1},\llbracket x_{2]},u\rrbracket\rrbracket\rrbracket$. Then \eqref{eq:jacobitheta} implies
\begin{align}
-\llbracket\Theta,\llbracket x_{[1},\llbracket x_{2]},u\rrbracket\rrbracket\rrbracket&=-\llbracket\llbracket\Theta,x_{[1}\rrbracket,\llbracket x_{2]},u\rrbracket\rrbracket + \llbracket x_{[1},\llbracket\Theta,\llbracket x_{2]},u\rrbracket\rrbracket\rrbracket\nonumber\\
&=-\llbracket\llbracket\llbracket\Theta,x_{[1}\rrbracket,x_{2]}\rrbracket,u\rrbracket-\llbracket x_{[2},\llbracket \llbracket\Theta,x_{1]}\rrbracket,u\rrbracket\rrbracket-\llbracket x_{[2},\llbracket\Theta,\llbracket x_{1]},u\rrbracket\rrbracket\rrbracket\nonumber\\
&=-\llbracket[x_1,x_2],u\rrbracket-2\llbracket x_{[2},\llbracket \Theta,\llbracket x_{1]},u\rrbracket\rrbracket\rrbracket-\llbracket x_{[2},\llbracket x_{1]},\llbracket\Theta, u\rrbracket\rrbracket\rrbracket,
\end{align}
which, by \eqref{eq:defbullet} and \eqref{eq:defdiff}, gives back $[x_1,x_2]\bullet u+2x_{[2}\bullet\mathfrak{D}\big(x_{1]}\bullet u\big)+x_{[2}\bullet \big(x_{1]}\bullet \mathfrak{D}(u)\big)$, that is, the right hand side of Axiom 4.

\textbf{Axiom 5:} 
Recalling that $|u|=\ell(u)-1$, the left hand side is equal to $(-1)^{\ell(u)+1}$ times
\begin{align}\label{eq:jacobitheta2}
\llbracket \Theta,\llbracket u,v\rrbracket\rrbracket-\llbracket \llbracket \Theta,u\rrbracket,v\rrbracket-(-1)^{\ell(u)}\llbracket u,\llbracket \Theta,v\rrbracket\rrbracket,
\end{align}
which is zero according to (\ref{eq:jacobitheta}).

\textbf{Axiom 6:} Since $\ell(a\bullet b)=\ell(a)+\ell(b)-2$, the left hand side is equal to $(-1)^{\ell(b)+1}$ times
\begin{align}
\llbracket a,\llbracket b,c\rrbracket\rrbracket-\llbracket\llbracket a,b\rrbracket,c\rrbracket+(-1)^{\ell(b)\ell(c)}\llbracket\llbracket a,c\rrbracket,b\rrbracket,
\end{align}
which is zero according to the Jacobi identity (\ref{jacobi}).

\textbf{Axiom 7:}
The nilpotency of $\mathfrak{D}$ follows from the nilpotency of $\Theta$ (by the Jacobi identity),
$\mathfrak{D}^2=\frac{1}{2}\llbracket\llbracket\Theta,\Theta\rrbracket,-\rrbracket=0$, hence turning $(X,\mathfrak{D})$ into a chain complex.

The identities \eqref{Leibnizhigher2} and \eqref{Leibnizhigher} characterizing the generalized Lie derivative can also be unified in the dgLa setting. For any $x\in X_0$ we set
\begin{align}
\mathcal{L}_x\equiv \llbracket\llbracket\Theta,x\rrbracket,\,-\,\rrbracket. \label{definitionliederivative}
\end{align}
Then \eqref{Leibnizhigher2} is nothing but \eqref{eq:defcircle}, whereas \eqref{Leibnizhigher} is implied by the Jacobi identity \eqref{eq:jacobitheta}.
Covariance and closure conditions, which are proven from the Axioms in \cite{Bonezzi:2019ygf}, follow here from the Jacobi identity on $T$. This construction is functorial, for every morphism of dgLa $\phi:T\to T'$ canonically induces a morphism of infinity-Leibniz algebras $f_\phi:X\to X'$. Let $F:\textbf{dgLa}_{\geq0}\to \infty\textbf{-enLeib}$ be the functor that associates an infinity-enhanced Leibniz algebra $X$ to any non-negatively graded dgLa $T$.

Conversely, given an infinity-enhanced Leibniz algebra $X=X_0\oplus X_1\oplus \cdots=X_0\oplus \overline{X}$, we will now show that one can canonically define a dgLa structure on some non-negatively graded vector space $T=T_0\oplus T_1\oplus\cdots=T_0\oplus \overline{T}$. First, let us shift the degree of all subspaces by $+1$, \emph{i.e.}, $\overline{T}=s^{-1}X$. That is to say, $T_{i}=X_{i-1}$ for every $i\geq1$, and in particular $T_1=X_0=V$. Let us set
\begin{align}
\llbracket a,b\rrbracket &\equiv (-1)^{|a|+1}a\bullet b,\label{eq:defbullet0}\\
\partial&\equiv -\mathfrak{D} \label{eq:defdiff0},
\end{align}
for $a,b \in X$ and $\partial$ acting on $s^{-1}\overline X$. The nilpotency of the operator $\mathfrak{D}$ implies that $\partial$ acts as  a differential on $\overline{T}$. 
As it is, $\big(T_1\oplus T_2\oplus\cdots,\llbracket\,.\,,.\rrbracket,\partial\big)$ is not a dgLa, since the Leibniz rule might not be satisfied on 
a bracket of two elements in $T_1$, since $\partial(T_1)=-\mathfrak{D}(X_0)=0$. Therefore, at degree 0 we let $T_0$ be the Lie algebra generated by the generalized Lie derivatives\footnote{By \eqref{Leibnizhigher2}, we see that the generalized Lie derivatives are endomorphisms of $\mathfrak{gl}(V)$, so $T_0$ can be seen as a sub-Lie-algebra of $\mathfrak{gl}(V)$. Moreover, by \eqref{eq:embeddingtensor}, we have that $\mathcal{L}_x(y)=x\circ y=x_L(y)$ for every $x,y\in V$. Hence, $\mathcal{L}_x$ and $x_L$ define the same endomorphism on $V$, however it could well happen that for some $x\in\mathcal{Z}$, the right hand side of \eqref{Leibnizhigher} is not vanishing, implying that $T_0$ may be bigger than $\bigslant{V}{\mathcal{Z}}$.  This is consistent with the fact that the map $x\longmapsto \mathcal{L}_x$ defines an embedding tensor $\Theta:V\to\mathfrak{gl}(V)$. Then, by \eqref{inclusion}, we obtain again that $T_0=\mathrm{Im}(\Theta)\simeq\bigslant{V}{\mathrm{Ker}(\Theta)}$ may have a bigger dimension than $\bigslant{V}{\mathcal{Z}}$.}, where the Lie bracket is given by the closure condition \eqref{closurecondition}. 
The bracket between $T_0$ and any other $T_i$, for $i\geq1$, is induced by the $T_0$-module structure on $T_i$, as defined in \eqref{Leibnizhigher2} and \eqref{Leibnizhigher}:
\begin{align}
\llbracket \mathcal{L}_x,a\rrbracket\equiv\mathcal{L}_x(a) 
\end{align}
for any $x\in T_1$ and $a\in \overline{T}$.
We enforce the skew-symmetry by setting $\llbracket a,\mathcal{L}_x\rrbracket =-\mathcal{L}_x(a)$. 
Finally, the differential $\partial$ can be extended to $T_1$ by setting
\begin{align}
\partial(x)\equiv\mathcal{L}_x
\end{align}
 for any $x\in T_1$.

Then, the skew-symmetry of the graded bracket $\llbracket\,.\,,.\,\rrbracket$ on $\overline{T}$ (see \eqref{gradedantisym}) is a  consequence of
\eqref{eq:definitionbullet}. Axiom 6 implies that the bracket satisfies the Jacobi identity \eqref{jacobi} on $\overline{T}$. Covariance and closure conditions imply that the graded bracket $\llbracket\,.\,,.\,\rrbracket$ satisfies the Jacobi identity also when any one or two elements in $T_0$ is involved. The Jacobi identity on $T_0$ is automatically satisfied because $T_0$ is a Lie algebra.
Proving the Leibniz rule \eqref{eq:derivationpartial} is a bit more involved. Axiom 2 is the Leibniz rule for two elements of $T_1$. Axiom 5 is the Leibniz rule for two elements of $\overline{T}$. The Leibniz rule for one element $x$ of $T_1$ and any other element $u$ of $\overline{T}$ is induced by \eqref{Leibnizhigher} and the definition of the differential
$\partial:T_1\to T_0,\,x\mapsto \mathcal{L}_x$. The last Leibniz identities we need to check are those involving at least one element of $T_0$. The Leibniz rule for an element of $T_0$ and any element in $\overline{T}$ is induced by the covariance of $\mathfrak{D}$. The Leibniz rule for an element of $T_0$ and any element of $T_1$ is induced by the covariance of $\circ$.

Hence, we have proven that $(T,\llbracket\,.\,,.\,\rrbracket,\partial)$ is a differential graded Lie algebra that is canonically induced from the data defining the infinity-enhanced Leibniz algebra.  Also this construction is functorial, for any morphism of infinity-enhanced Leibniz algebras $f:X\to X'$ canonically induces a morphism of dgLa $\phi_f:T\to T'$. The correspondence is obvious on $\overline{T}$, and on $T_0=\{\mathcal{L}_x\ | \ x\in X_0\}$, the obvious definition is $\phi_f(\mathcal{L}_x)=\mathcal{L}_{f(x)}$. Then by \eqref{Leibnizhigher2}--\eqref{Leibnizhigher} and the fact that $f$ is a morphism in the category of infinity-enhanced Leibniz algebras, $\phi_f\big(\llbracket x, a\rrbracket\big)=\llbracket \phi_f(x),\phi_f(a)\rrbracket$ for every $x\in T_0$ and $a\in T$. This proves that $\phi_f:T\to T'$ is a well-defined morphism in the category of dgLa.  Let $G:\infty\textbf{-enLeib}\to\textbf{dgLa}_{\geq0}$ be the functor that associates a dgLa $T$ to any infinity-enhanced Leibniz algebra $X$ by the construction above.

Moreover, notice that according to the discussion in Section \ref{sectiondgLa}, the dgLa $T$ can then be canonically extended to a dgLa
$T_{-1} \oplus T_0 \oplus T_1 \oplus \cdots$, where $T_{-1}$ is a one-dimensional subspace spanned by the embedding tensor $\Theta$. To conclude, since any non-negatively graded dgLa $T$ induces a Leibniz product on $T_1$ (see Section \ref{sectiondgLa}), and a compatible infinity-enhanced Leibniz algebra (this section), and since any Leibniz algebra $V$ gives rise to a non-negatively graded dgLa (see Section \ref{sectionTHA}) whose induced Leibniz product coincides with the original one, we have an explicit construction of an infinity-enhanced Leibniz algebra 
starting from only the
Leibniz algebra $V$ (see the diagram in the introduction).

The question of the functoriality of both constructions is worth some attention. One notices that for any infinity-enhanced Leibniz algebra $X$, we have $F(G(X))=X$, but not every non-negatively graded dgLa satisfies $G(F(T))=T$ -- and in general it is not true. This is due to the fact that $F$ forgets the degree 0 part of $T$. Hence in full generality $F$ and $G$ do not define an equivalence of categories. However, the image of the functor $G$ defines a full subcategory of $\textbf{dgLa}_{\geq0}$, on which the composition of $G$ and $F$ is the identity, thus defining an equivalence of categories with $\infty\textbf{-enLeib}$. Then, non-negatively graded dgLa provide a natural way of coding infinity-enhanced Leibniz algebras, without any loss of data\footnote{There are actually many more non-negatively graded dgLa than there are infinity-enhanced Leibniz algebras, because the degree 0 part of such a dgLa $T$ is forgotten by the functor $G$. This opens the question of the physical information that may be contained in $T_0$, and that cannot be captured by the associated infinity-enhanced Leibniz algebra $G(T)$.}. 
One of the advantages with the dgLa structure is that it
gives rise to an $L_\infty$-algebra structure in a canonical and explicit way, as we will see in the next section.

\section{The $L_\infty$-algebra induced by the dgLa}\label{sectiongetzler}

The link between tensor hierarchies and $L_\infty$-algebras goes back to \cite{Palmer:2013pka,Lavau:2014iva}, where an $L_3$-algebra structure was induced from the field strengths of a superconformal $(1,0)$ model in six dimensions \cite{Samtleben:2011fj}. We will come back to this example in the end of this section.
Our aim is to show how the dgLa structure of the tensor hierarchy gives rise to a canonical
$L_\infty$-algebra structure.
This result is a consequence of a theorem by Getzler \cite{Getzler:2010}, which in turn is a special
case of a more general theorem by Fiorenza and Manetti \cite{FiorenzaManetti}. 
We then show that the first few brackets of this $L_\infty$-algebra are the ones defined in \cite{Bonezzi:2019ygf}.

This section also implies a more general result for Leibniz algebras: the skew-symmetric part $[\,.\,,.\,]$ of the Leibniz product is in general not a Lie bracket since it does not satisfy the Jacobi identity, see \eqref{jacobiator0}. The material presented in this section implies however that this bracket actually fits precisely in the $L_\infty$-algebra structure that is defined by Getzler's theorem from the dgLa induced by the Leibniz algebra. This means in particular that the skew-symmetric part of the Leibniz product of any Leibniz algebra can be canonically \emph{lifted} to an $L_\infty$-algebra structure (such that if the Leibniz algebra is a Lie algebra, then this lift is trivial in the sense that the $L_\infty$-algebra reduces to an Lie 1-algebra which is this Lie algebra itself). This result {differs from the one presented in \cite{Kotov:2018vcz}, where the $L_\infty$-algebra constructed from the Leibniz algebra $V$ does not coincide with $V$ itself when $V$ is a Lie algebra.}

\subsection{$L_\infty$-algebras and Getzler's theorem}\label{getzlertheorem}

The notion of an $L_\infty$-algebra generalizes the notion of a differential graded Lie algebra by weakening the Jacobi identity, and allowing it to be satisfied only \emph{up to homotopy} \cite{Lada:1992wc,Marklada}.
The precise definition of $L_\infty$-algebras can be found in many papers, see for example \cite{Hohm:2017pnh}, 
and we will only recall the basics here. We emphasize that there are two different conventions, with graded symmetric and graded skew-symmetric brackets, respectively.
We will use the usual skew-symmetric convention here, as it it more adapted to our setting. The symmetric convention for $L_\infty$-algebras is given in \cite{Mehta}, and the correspondence between the skew-symmetric and the symmetric brackets is rigorously defined in Remark 1.1 of \cite{FiorenzaManetti}, see \eqref{translation}. A discussion about 
this correspondence can also be found in \cite{Cederwall:2018aab}.

An $L_\infty$-algebra is a $\mathbb{Z}$-graded vector space $X=\bigoplus_{i\in\mathbb{Z}}X_i$ that is equipped with a family $(l_k)_{k\geq1}$ of degree $(k-2)$ graded skew-symmetric $k$-multilinear `brackets'
that satisfy the \emph{higher Jacobi identities}, written symbolically as\footnote{The rigorous formula is \begin{equation}\label{superjacobi}
\sum_{i+j=n+1}\ (-1)^{i(j-1)} \sum_{\sigma\in \textrm{Un}(i,n-i)}\epsilon^\sigma_{x_1,\ldots,x_n}\,l_{j}\big(l_i(x_{\sigma(1)},\ldots,x_{\sigma(i)}),x_{\sigma(i+1)},\ldots,x_{\sigma(n)}\big)=0
\end{equation} where $\textrm{Un}(i,n-i)$ is the set of $(i,n-i)$-unshuffles and where $\epsilon^\sigma_{x_1,\ldots,x_n}$ is the sign induced by the permutation of the elements $x_1,\ldots, x_n$ in the exterior algebra of $X$, {\it i.e.,} $x_1\wedge \ldots x_n=\epsilon^\sigma_{x_1,\ldots,x_n} x_{\sigma(1)}\wedge \ldots \wedge x_{\sigma(n)}$.}
\begin{align}\label{higherJac}
\sum_{i+j=n+1}^n(-1)^{i(j-1)}l_j\circ l_i=0.
\end{align}
In particular, $l_1$ is a differential on $X$, and it is a derivation of the 2-bracket $l_2$.
We say that $X$ is a Lie $n$-algebra if it is non-negatively graded and concentrated in degrees $0,\ldots,n-1$: $X=X_0\oplus\cdots \oplus X_{n-1}$. In particular, this implies that a Lie algebra is a Lie 1-algebra.

A theorem by Getzler \cite{Getzler:2010} shows that a differential graded Lie algebra structure on $T=T_0\oplus T_1\oplus T_2\oplus\cdots=T_0\oplus \overline{T}$ canonically induces a $L_\infty$-algebra structure on $X=s\,\overline{T}$.
In \cite{Getzler:2010} formulas for the brackets of all orders are given, but there appears to be an error in the formulas
that can be corrected by reversing the sign of the higher odd brackets. It is also mentioned in 
\cite{Getzler:2010} that the theorem is a special case of a more general result
already found by Fiorenza and Manetti \cite{FiorenzaManetti}\footnote{See for example Remark 5.4 in \cite{FiorenzaManetti} and there replace the differential $\partial$ with the map $D$ defined in \cite{Getzler:2010}. Unfortunately, they also made a mistake in this application: for $n\geq2$ the right hand side of the formula computing the $n+1$ bracket should inherit a minus sign.}. To pass from the symmetric convention for $L_\infty$-algebras used in \cite{Getzler:2010} to the usual skew-symmetric convention, one uses the rigorous formula given in Remark 1.1 of \cite{FiorenzaManetti}:
\begin{align}\label{translation}
l_n(a_1,\ldots, a_n)=(-1)^{n(n-2)+\sum_{i=1}^{n-1}(n-i)|a_{i}|}\big\{a_1,\ldots, a_n\big\}_n
\end{align}
where $\{\ldots\}_n$ is the $n$-bracket in the symmetric convention, and where as usual $|a|$ is the degree of $a$ when seen as an element of $X$.
In particular, the differential and the odd brackets inherit an additional minus sign when performing this transformation.

 Let us now turn to the result of Getzler, translated to the convention of
 skew-symmetric brackets and with reversed signs of the odd brackets of order 3 and higher, correcting the sign error in the original statement of the theorem.
 
 \bigskip
 
 \begin{theoreme*}
 A differential graded Lie algebra $(T,\llbracket\,.\,,.\,\rrbracket,\partial)$
 gives rise to the following $L_\infty$-algebra structure on $X=s\,\overline{T}$:
\begin{enumerate}
\item the 1-bracket is $l_1\equiv-\partial$ on $\overline{X}$\footnote{Since Getzler's 1-bracket is $\{a\}=\partial(a)$, we pick up a minus sign from
\eqref{translation}.} and $0$ on $X_0$, where $X=X_0\oplus \overline{X}$;
\item the $2$-bracket is defined by
\begin{align}
l_2(a,b)\equiv\frac{(-1)^{|a|}}{2}\big(\llbracket D(a),b\rrbracket+(-1)^{\ell(a)\ell(b)}\llbracket D(b),a\rrbracket\big),\label{getzler1}
\end{align}
where $D:T\to T$ is the operator that is equal to $\partial$ on $T_1$, and $0$ in any other degree;
\item the $k$-bracket for $k\geq3$ is given by
\begin{align}
l_k(a_1,\ldots, a_{k})\equiv\beta_k\sum_{\sigma\in S_k}\chi^\sigma_{a_1,\ldots,a_k}\llbracket\llbracket\ldots\llbracket\llbracket D(a_{\sigma(1)}),a_{\sigma(2)}\rrbracket,a_{\sigma(3)}\rrbracket\ldots\rrbracket a_{\sigma(k)}\rrbracket,\label{getzler2}
\end{align}
where $\beta_k=(-1)^{\sum_{i=1}^k(k-i)|a_i|}\frac{B_{k-1}}{(k-1)!}$, with $B_{k-1}$ being the $(k-1)$-th Bernoulli number\footnote{Getzler's definition of higher brackets does not satisfy the definition of a $L_\infty$-algebra, but it does if we reverse the sign of the odd brackets of order 3 and higher (even brackets of order 4 and higher are vanishing anyway since $B_3=B_5=\cdots=0$). Hence, when passing to the skew-symmetric convention, the odd  brackets inherit a plus sign, because the sign brought by the translation from the symmetric convention to the skew-symmetric convention (see \eqref{translation}) cancels the sign that we added to correct Getzler's formula for odd brackets of order 3 and higher.} and the sign $\chi^\sigma_{a_1,\ldots,a_k}$ is the Koszul sign of the permutation $\sigma$ with respect to the degree in $T$, that is: $a\vee b=(-1)^{\ell(a)\ell(b)}b\vee a$ so that $\chi^{(1\, 2)}_{a,b}=(-1)^{\ell(a)\ell(b)}$. Since $B_3=B_5=\cdots=0$ there is no $k$-bracket for $k$ even and greater than 3.
\end{enumerate}
\end{theoreme*}
The occurrence of Bernoulli numbers in this context was first noted by Bering \cite{Bering:2006eb}, and they also show up in a similar way in the $L_\infty$ algebra
encoding the gauge structure of generalised diffeomorphisms in extended geometry \cite{Cederwall:2018aab}.

\subsection{Comparing the $L_\infty$-algebra structures}

Using Getzler's theorem, let us compute the $L_\infty$-algebra $(X,l_k)$ that is associated to the differential graded Lie algebra $(T,\llbracket\,.\,,.\,\rrbracket, \llbracket\Theta,-\rrbracket)$ induced by a Leibniz algebra $V$ (see Section \ref{sectionTHA}). First, notice that the underlying graded vector space defined in Getzler's theorem is the same as the one used in \cite{Bonezzi:2019ygf}, that is: $X=s\,\overline{T}$. We will now show that the $L_\infty$-algebra defined on $X$ by using Getzler's theorem gives back all brackets defined from the infinity-enhanced Leibniz algebra structure on $X$ in \cite{Bonezzi:2019ygf}.

First, the 1-bracket on $X$ is $l_1=-\llbracket\Theta,-\rrbracket=\mathfrak{D}$, which is the same  differential as in \cite{Bonezzi:2019ygf}.
Following Getzler, we define the operator $D:T\to T$ as $D=\llbracket\Theta,-\rrbracket$ on $T_1=V$, and 0 elsewhere. Then, the 2-bracket of two elements $x,y$ of $X_0=V$, which are considered as having degree 0 in $X$ so that $|x|=|y|=0$, is given by \eqref{getzler1}:
\begin{align}
l_2(x,y)=\frac{1}{2}\Big(\llbracket\llbracket\Theta,x\rrbracket,y\rrbracket -\llbracket \llbracket\Theta,y\rrbracket,x\rrbracket \Big)=\frac{1}{2}\big(x\circ y -y\circ x\big)=[x,y],
\end{align}
where we have used \eqref{eq:moduleV}. Hence, the 2-bracket of two elements of $X_0$ is precisely the skew-symmetric part of the Leibniz product. Now, let us compute the bracket of an element $x\in X_0=V$ and another element $u\in \overline{X}$. Since $|u|>0$, then $D(u)=0$ so that only the first term of \eqref{getzler1} appears:
\begin{align}
l_2(x,u)=\frac{1}{2}\llbracket\llbracket\Theta,x\rrbracket,u\rrbracket=\frac{1}{2}\mathcal{L}_x(u),
\end{align}
where we have used \eqref{definitionliederivative}. The bracket with two elements in $\overline{X}$ vanishes because the operator $D$ vanishes on $s^{-1}\overline{X}$. Thus, we see that
the 2-bracket defined by Getzler's theorem from the dgLa induced by a Leibniz algebra coincides with the one of \cite{Bonezzi:2019ygf}.

For the 3-bracket, the formula \eqref{getzler2} gives:
\begin{align}
l_3(a,b,c)&=\frac{(-1)^{2|a|+|b|}}{12}\Big(\llbracket\llbracket D(a),b\rrbracket,c\rrbracket+(-1)^{\ell(b)\ell(a)}\llbracket\llbracket D(b),a\rrbracket,c\rrbracket\\
&\hspace{1.4cm}+(-1)^{\ell(a)(\ell(b)+\ell(c))}\llbracket\llbracket D(b),c\rrbracket,a\rrbracket+(-1)^{\ell(a)(\ell(b)+\ell(c))+\ell(b)\ell(c)}\llbracket\llbracket D(c),b\rrbracket,a\rrbracket\nonumber\\
&\hspace{1.8cm}+(-1)^{(\ell(a)+\ell(b))\ell(c)}\llbracket\llbracket D(c),a\rrbracket,b\rrbracket+(-1)^{\ell(b)\ell(c)}\llbracket\llbracket D(a),c\rrbracket,b\rrbracket\Big).\nonumber
\end{align}
For three elements $x_1,x_2,x_3$ of $X_0=V$, we have $|x_i|=0=\ell(x_i)-1$, so that we obtain
\begin{align}
l_3(x_1,x_2,x_3)=\frac{1}{2}\llbracket x_{[1}\circ x_2,x_{3]}\rrbracket=-\frac{1}{2} (x_{[1}\circ x_2)\bullet x_{3]}.
\end{align}
Including one element $u_n\in X_n$ for some $n\geq1$, so that $|u_n|=n=\ell(n)-1$, we get
\begin{align}
l_3(x_1,x_2,u_n)&=\frac{1}{6}\Big(\llbracket [x_{1},x_2],u_n\rrbracket+(-1)^n\llbracket \mathcal{L}_{x_{[2}}(u_n),x_{1]}\rrbracket\Big)\\
&=-\frac{1}{6}[x_{1},x_2]\bullet u_n-\frac{1}{6}\mathcal{L}_{x_{[2}}(u_n)\bullet x_{1]}.
\end{align}
Finally, for the 3-bracket of one element $x\in X_0=V$ and two other elements $u_n\in X_n$ and $u_m\in X_m$ ($m,n\geq1$), we have
\begin{align}
l_3(x,u_n,u_m)&=\frac{(-1)^n}{12}\Big(\llbracket\mathcal{L}_{x}(u_n),u_m\rrbracket+(-1)^{\ell(n)\ell(m)}\llbracket\mathcal{L}_{x}(u_m),u_n\rrbracket\Big)\\
&=-\frac{(-1)^{\ell(n)}}{12}\Big(\llbracket\mathcal{L}_{x}(u_n),u_m\rrbracket-\llbracket u_n,\mathcal{L}_{x}(u_m)\rrbracket\Big)\\
&=-\frac{1}{12}\Big(\mathcal{L}_{x}(u_n)\bullet u_m-u_n\bullet\mathcal{L}_{x}(u_m)\Big)\\
&=\frac{1}{12}\Big(u_n\bullet\mathcal{L}_{x}(u_m)-(-1)^{|n||m|}u_m\bullet \mathcal{L}_x(u_n)\Big),
\end{align}
and with only elements in $\overline X$, the 3-bracket vanishes in the same way as the 2-bracket.

We observe that the 3-brackets are exactly the ones defined in \cite{Bonezzi:2019ygf}. 
Given that Bernoulli numbers $B_{k-1}$ vanish for $k$ even and greater than 3, all even brackets vanish, as was found for the 4-bracket in \cite{Bonezzi:2019ygf}. The $L_\infty$-algebra structure on $X$ obtained from Getzler's theorem, and the one defined on $X$ \cite{Bonezzi:2019ygf} thus coincide up to that order. However, Getzler's theorem provides us with the precise formulas for higher brackets, that were not given in \cite{Bonezzi:2019ygf}.

This result is mathematically very deep because it says that given any Leibniz algebra $V$, one can always find a (non-negatively graded) $L_\infty$-algebra that `lifts' the skew-symmetric part of the Leibniz product in a non-trivial way (recall that this bracket does not satisfy the Jacobi identity, see \eqref{jacobiator0}). More precisely, the vector space at degree 0 is $V$ and the 2-bracket of the $L_\infty$-algebra at degree 0 is $[\,.\,,.\,]$. This $L_\infty$-algebra has the particularity that if $V$ is a  Lie algebra, {\it i.e.}, if the symmetric part of the Leibniz product is zero, then $X=V$. This can be explained by the construction of the dgLa induced by the Leibniz algebra $V$ in Section~\ref{sectionTHA} and with more details in \cite{Lavau:2017tvi}: if $V$ is a Lie algebra, the dgLa does not have any space of degree higher than 1. Hence, that is why the $L_\infty$-algebra $X$ defined from a Leibniz algebra $V$ by applying Getzler's theorem to the dgLa induced by $V$ can be called the \emph{$L_\infty$-extension of the Leibniz algebra $V$}. We postpone to further research the study of such algebraic structures.

\subsection{Example: the $(1,0)$ superconformal model in six dimensions}

The first levels of the tensor hierarchy appearing in the $(1,0)$ superconformal model in six dimensions have been given in \cite{Samtleben:2011fj}. Its mathematical aspects were investigated in \cite{Palmer:2013pka,Lavau:2014iva}. The algebra of global symmetries of this model is $\mathfrak{g}\equiv\mathfrak{e}_{5(5)}=\mathfrak{so}(5,5)$ \cite{Trigiante:2016mnt}. The model involves a set of $p$-forms (for $p=1,2,3,\ldots,6$) taking values, respectively, in the following $\mathfrak{g}$-modules: $X_0=\textbf{16}$, $X_1=\textbf{10}$, $X_2=\overline{\textbf{16}}$, $X_3=\textbf{45}$, $X_4=\overline{\textbf{144}}$ and $X_5=\textbf{10}\oplus\overline{\textbf{126}}\oplus\textbf{320}$ \cite{Trigiante:2016mnt}.
The embedding tensor $\Theta:X_0\to\mathfrak{g}$ is considered as a \emph{spurionic object}, \emph{i.e.}, that is only fixed at the end of the computations, thus immediately fixing any tensors appearing in the hierarchy. In the (1,0) superconformal model, the rank of $\Theta$ is constant, so that we can formally set $\mathfrak{h}\equiv\mathrm{Im}(\Theta)\subset \mathfrak{g}$, without further addressing the content of this Lie subalgebra.

The beginning of the hierarchy  is governed by a set of constants $h^a_I$, $g^{It}$, 
$f_{ab}^c = -f_{ba}^c$,  $d^I_{ab} = d^I_{ba}$, $b^{}_{Ita}$, 
subject to the following relations:
\begin{align}
	2\big(d^{J}_{c(a}d^{I}_{b)s}-d^{I}_{cs}d^{J}_{ab}\big)h^{s}_{J}			&=2f_{c(a}{}^{s}d^{I}_{b)s}-b^{}_{Jsc}d^{J}_{ab}g^{Is},\label{lol1}\\
	\big(d^{J}_{rs}b^{}_{Iut}+d^{J}_{rt}b^{}_{Isu}+2d^{K}_{ru}b^{}_{Kst}\delta^{J}_{I}\big)h^{u}_{J}&=f_{rs}{}^{u}b^{}_{Iut}+f_{rt}{}^{u}b^{}_{Isu}+g^{Ju}b^{}_{Iur}b^{}_{Jst}\label{lol2},\\
	f_{[ab}{}^{r}f_{c]r}{}^{s}-\frac{1}{3}h_{I}^{s}d^{I}_{r[a}f_{bc]}{}^{r}&=0\label{lol3},\\
	h^{a}_{I}g^{It}													&=0\label{lol4},\\
	f_{rb}{}^{a}h_{I}^{r}-d^{J}_{rb}h^{a}_{J}h^{r}_{I}				&=0\label{lol5},\\
	g^{Js}h^{r}_{I}b^{}_{Ksr}-2h_{K}^{s}h_{I}^{r}d^{J}_{rs}				&=0\label{lol6},\\
	-f_{rt}{}^{s}g^{It}+d^{J}_{rt}h^{s}_{J}g^{It}-g^{It}g^{Js}b^{}_{Jtr}	&=0\label{lol7},\\
	b^{}_{Jt(a}d^{J}_{bc)}												&=0\label{lol8}.
\end{align}	

Following \eqref{condition3}, the  $\mathfrak{g}$-module $X_0=\textbf{16}$ can be equipped with a Leibniz algebra structure $\circ$ that is defined between any two basis elements $e_a,e_b\in X_0$ by
\begin{equation}\label{conformaleibniz}
e_{a}\circ e_b\equiv-X_{ab}{}^c\, e_c.
\end{equation}
We call $X_{ab}{}^c=-f_{ab}{}^c+d^I_{ab}h^c_I$ the \emph{structure constants} of the Leibniz algebra. 
The symmetric and skew-symmetric brackets are then defined by
\begin{equation}
[e_a,e_b]=f_{ab}{}^c\,e_c\hspace{1cm}\text{and}\hspace{1cm}\{e_a,e_b\}=-d^I_{ab}h_I^c\, e_c.
\end{equation}
Then, the skew-symmetric part of the Leibniz product does not satisfy the Jacobi identity, but rather \eqref{jacobiator0}, which is precisely \eqref{lol3}.

The second $\mathfrak{g}$-module $X_1=\textbf{10}$ is a sub-representation of $S^2(X_0)$ that is fixed by the representation constraint, and that turns out to be the smallest $\mathfrak{g}$-submodule of $S^2(X_0)$ through which the symmetric bracket factors (see Section \ref{sectionTHA}). We set  $\{e_I\}$ to be a basis of $X_1$, and following the notations in \cite{Samtleben:2011fj}, we label these elements by capital letters of the middle of the alphabet, not to confuse them with the generators of $X_0$. The action of $\mathfrak{h}$ on $X_1$ is defined by
\begin{equation}\label{return2}
\rho_{\Theta(e_a)}(e_I)\equiv-X_{aI}{}^J\, e_J,
\end{equation}
where $X_{aI}{}^J=2h_{I}^cd^J_{ac}-g^{Js}b^{}_{Isa}$, and where $e_a$ is a basis element of $X_0$.
Going further up, we reach the $\mathfrak{g}$-module $X_2=\overline{\textbf{16}}$ in which 3-form fields take values. In the $(1,0)$ superconformal model, the 3-forms are dual to the 1-forms, that is why we use latin letters of the end of the alphabet as labels for basis elements $\{e^t\}$ of $X_2$. 
By duality, the action of $\mathfrak{h}$ on a generator $e^t$ is defined by
\begin{equation}\label{return3}
\rho_{\Theta(e_a)}(e^t)\equiv X_{as}{}^t\,e^s,
\end{equation}
where $X_{as}{}^t=-f_{as}{}^t+d^I_{as}h^t_I$, and where $e_a$ is a basis element of $X_0$. 

As explained in \cite{Samtleben:2011fj}, the hierarchy of $p$-forms (for $p=1,2,3$) can be extended one step further by adding a set of 4-forms that take values in the $\mathfrak{g}$-module $X_3=\textbf{45}$.
 Three new tensors $k_t^\alpha, c^{}_{\alpha IJ}=-c^{}_{\alpha JI}$ and $c^t_{\alpha a}$ have to be introduced  so that this extension is consistent. They obey a set of additional conditions:
\begin{align}
	g^{Kt}k_{t}^{\alpha}												&=0\label{lol9},\\
	4d^J_{ab} c^{}_{\alpha IJ} - b^{}_{Ita} c^t_{\alpha b} - b^{}_{Itb} c^t_{\alpha a}
	&=0\label{lol10},\\
	k_t^\alpha c^{}_{\alpha IJ} -h^a_{[I}b^{}_{J]ta} 
	&=0\label{lol11},\\
	k_t^\alpha c^s_{\alpha a} - f_{ta}{}^s + b^{}_{Jta}g^{Js} - d^J_{ta}h^s_J
	&=0\label{lol12}.
\end{align}
The tensor hierarchy is not investigated higher than this level, so we will only consider the structures that are induced by \eqref{lol1}--\eqref{lol8} and 
\eqref{lol9}--\eqref{lol12}, keeping in mind that the hierarchy formally goes higher up.

The construction of the tensor hierarchy presented in Section \ref{sectionTHA} and with more details in \cite{Lavau:2017tvi} gives the following brackets on the tensor hierarchy $T=\mathfrak{h}\oplus s^{-1}X_0\oplus s^{-1}X_1\oplus s^{-1}X_2\oplus \cdots$: 
 \begin{align}
	\llbracket X_{a},X_{b}\rrbracket	&=	-2\,d^{I}_{ab}\,X_{I}\label{brak1},\\
	\llbracket X_{a},X_{I}\rrbracket	&=	-b^{}_{Ita}\,X^{t}\label{brak2},\\
	\llbracket X_{I},X_{J}\rrbracket	&=	-2\,c^{}_{\alpha IJ}\,X^{\alpha}\label{brak3},\\
	\llbracket X_{a},X^{t}\rrbracket	&=	c_{\alpha a}^{t}\,X^{\alpha}\label{brak4}
\end{align}
where $X_a=s^{-1}e_a$, $X_I=s^{-1}e_I$ and $X^t=s^{-1}e^t$ (and similarly for $X_b$ and $X_J$).
Notice that elements of $T_1=s^{-1}X_0$ have degree $+1$, elements of $T_2=s^{-1}X_1$ have degree $+2$, elements of $T_3=s^{-1}X_2$ have degree $+3$, \emph{etc.} so that the brackets are indeed consistent with the symmetries of the tensors aforementioned. The bracket of $\mathfrak{h}$ and any other element if the hierarchy is given by the action of $\mathfrak{h}$:
\begin{align}
\llbracket g,a\rrbracket=\rho_g(a)
\end{align}
for any $g\in \mathfrak{h}$ and $a\in T$. Skew-symmetry
is enforced by setting $\llbracket a,g\rrbracket=-\rho_g(a)$. To complete the differential graded Lie algebra structure  on $T$, the differential $\partial$ is given by:
\begin{align}
\partial(X_a)&=\Theta(X_a),\\
\partial(X_I)&=h_I^a\,X_a,\\
\partial(X^t)&=-g^{It}\,X_I,\\
\partial(X^\alpha)&=k_t^\alpha\,X^t.
\end{align}
Notice that it is the opposite sign convention that has been chosen in Section 4.3 in \cite{Lavau:2017tvi}.

One can check that the homological property $\partial^2=0$, the Jacobi identity \eqref{jacobi} and the Leibniz rule \eqref{eq:derivationpartial} are equivalent to \eqref{lol1}--\eqref{lol8} and \eqref{lol9}--\eqref{lol12}. For example, the Jacobi identity between elements $X_a, X_b\in T_1$ and $X_I\in T_2$ is:
\begin{align}
\llbracket X_a,\llbracket X_b, X_I\rrbracket \rrbracket-\llbracket \llbracket X_a, X_b\rrbracket , X_I\rrbracket &+\llbracket X_b,\llbracket X_a, X_i\rrbracket\rrbracket =  \nonumber\\
&\quad=2\llbracket X_{(a|},-b_{It|b)}X^t\rrbracket+2d^{J}_{ab}\llbracket X_J,X_I\rrbracket  \nonumber\\
&\quad=\Big(-2c_{\alpha(a|}^tb_{It|b)}-4d^J_{ab}c_{\alpha JI}\Big)\, X^\alpha,
\end{align}
which vanishes by skew-symmetry of the $I-J$ indices of the tensor $c_{\alpha JI}$, and by \eqref{lol10}. Another example of Jacobi identity is the one between an element $\Theta(X_a)$ of $T_0=\mathfrak{h}=\mathrm{Im}(\Theta)$ and two elements $X_b,X_c$ of $T_1$:
\begin{align}
&\llbracket\Theta(X_a),\llbracket X_b, X_c\rrbracket \rrbracket-\llbracket\llbracket \Theta(X_a),X_b\rrbracket, X_c\rrbracket-\llbracket\llbracket X_b, \Theta(X_a),X_c\rrbracket \rrbracket=\nonumber\\
&\hspace{3cm}=-2d^I_{bc}\llbracket\Theta(X_a),X_I\rrbracket+X_{ab}{}^r\llbracket X_r,X_c\rrbracket+X_{ac}{}^r\llbracket X_b,X_r\rrbracket\nonumber\\
&\hspace{3cm}=2\Big(d^I_{bc}X_{aI}{}^J-2X_{a(b}{}^rd_{c)r}^J\Big)\,X_J,
\end{align}
which vanishes by using the definition of $X_{ab}{}^r$ and $X_{aI}{}^J$, and by \eqref{lol1}.

Now let us turn to some Leibniz identities. We have
    \begin{align}
    	\partial\big(\llbracket X_{I},X_{J}\rrbracket\big)	-\llbracket\partial(X_I),X_J\rrbracket-\llbracket X_I,\partial(X_J)\rrbracket&=	-2\,c^{}_{\alpha IJ}k^\alpha_t X^t-2h^a_{[I}\llbracket X_a,X_{J]}\rrbracket\nonumber\\
	&=-2\Big(\,c^{}_{\alpha IJ}k^\alpha_t -h^a_{[I}b^{}_{J]ta}\Big)\, X^t,
    \end{align}
  which vanishes by \eqref{lol11}. Another example using $\Theta$ is
  \begin{align}
  \partial\big(\llbracket X_a,X^t\rrbracket\big)-\llbracket\partial(X_a),X^t\rrbracket+\llbracket X_a,\partial(X^t)\rrbracket&=c^t_{\alpha a}k_s^\alpha\,X^s-\llbracket\Theta(X_a),X^t\rrbracket-g^{It}\llbracket X_a,X_I\rrbracket\nonumber\\
  &=\Big(c^t_{\alpha a}k_s^\alpha-X_{as}{}^t+g^{It}b^{}_{Isa}\Big)\,X^s,
  \end{align}
    which vanishes by the definition of $X_{as}{}^t=-f_{as}{}^t+d_{as}^Kh^t_K$ and \eqref{lol12}.

Let us now give the $L_\infty$-algebra that is induced from this dgLa, following Getzler's theorem presented in Section \ref{getzlertheorem}. First, the $L_\infty$-algebra structure is defined on the graded vector space $X=s\,\overline{T}=X_0\oplus X_1\oplus\cdots$. We will use the same basis elements for $X$ as above, that is: $e_a, e_b, e_c, e_d, e_e$ are basis elements of $X_0$, $e_I, e_J, e_K$ are basis elements of $X_1$, $e^s, e^t$ are basis elements of $X_2$, \emph{etc.} Then, Item 1. of Getzler's theorem states that $l_1\equiv-\partial$, so that we have
\begin{equation}\label{eq:infinity1}
l_1(e_I)=-h_I^a\,e_a\hspace{0.1cm},\hspace{0.9cm}l_1(e^t)=g^{It}\,e_I\hspace{0.5cm}\text{and}\hspace{0.5cm}l_1(e^\alpha)=-k_t^\alpha\,e^t.
\end{equation}
The 2-bracket is defined as in \eqref{getzler1}:
\begin{align}
l_2(e_a,e_b)&=f_{ab}{}^ce_c,\\
l_2(e_a,e_I)&=-\frac{1}{2}X_{aI}{}^{J}e_J,\\
 l_2(e_a,e^t)&=\frac{1}{2}X_{as}{}^te^s.
\end{align}
If the 2-bracket does not involve at least one element from $X_0$ then it is vanishing. Following \eqref{getzler2}, the 3-bracket satisfies at lower degrees:
\begin{align}
l_3(e_a,e_b,e_c)&=-f_{[ab}{}^rd_{c]r}^I\,e_I,\\
l_3(e_a,e_b,e_I)&=-\frac{1}{6}\Big(f_{ab}{}^{c}b_{Itc}+X_{[a|I}{}^{K}b_{Kt|b]}\Big)\,e^t,\\
l_3(e_a,e_b,e^t)&=-\frac{1}{6}\Big(-f_{ab}{}^{c}c_{\alpha c}^t+X_{[a|s}{}^{t}c_{\alpha|b]}^s\Big)\,e^\alpha,\\
l_3(e_a,e_I,e_J)&=-\frac{1}{3}X_{a(I|}{}^Kc_{\alpha K|J)}\,e^\alpha.\label{eq:infinity2}
\end{align}
Obviously there are 2-brackets and 3-brackets involving elements of the spaces $X_3$, $X_4$ and $X_5$, but we only focus here on the beginning of the hierarchy, using only the tensors that were given in \cite{Samtleben:2011fj}. Thus, the only 5-bracket that is computable given the available tensors is the one involving five elements of $X_0$:
\begin{align}
l_5(e_{\underline{a}},e_{\underline{b}},e_{\underline{c}},e_{\underline{d}},e_{\underline{e}})=\frac{1}{3}f_{\underline{a}\underline{b}}{}^rd_{r\underline{c}}^Ib_{It\underline{d}}c_{\alpha\underline{e}}^t\,e^\alpha,\label{eq:infinity3}
\end{align}
where underlined indices are fully-antisymmetric.
There is only one 7-bracket $l_7:\wedge^7X_0\to X_5$, but since not all tensors were given in \cite{Samtleben:2011fj} to define the whole hierarchy, we cannot write it down. The 4- and 6-brackets are vanishing because the Bernoulli numbers $B_3$ and $B_5$ vanish.
As a final remark, since the hierarchy in the $(1,0)$ superconformal model stops at degree 5, we will actually obtain a Lie 6-algebra.

We propose now to give explicit computations of the higher Jacobi identities \eqref{higherJac} to prove that the above brackets indeed form an $L_\infty$-algebra. 
The first  (higher) Jacobi identities are those involving $l_1$ and $l_2$: $(l_1)^2=0$ and $l_1\circ l_2- l_2\circ l_1=0$. They can be checked rather easily so we turn directly to the first Jacobi identity involving three elements, that is:
\begin{align}
3\,l_2\big(l_2(e_{[a}, e_b),e_{c]}\big)+l_1\big(l_3(e_a,e_b,e_c)\big)=\Big(3f_{[ab|}{}^{r}f_{r|c]}{}^d+h^d_If_{[ab}{}^rd_{c]r}^I\Big)\,e_d,
\end{align}
which vanishes by the skew-symmetry of the lower indices of $f_{rc}{}^d$ and by \eqref{lol3}. The second (higher) Jacobiator that we can compute involves $e_a, e_b$ and $e_I$, and is defined by
\begin{align}
\mathrm{J}(e_a,e_b,e_I)&\equiv l_2\big(l_2(e_a, e_b),e_I\big)-2\,l_2\big(l_2(e_{[a|},e_I),e_{|b]}\big)\nonumber\\
&\quad\,+l_1\big(l_3(e_a,e_b,e_I)\big)+l_3\big(l_1(e_I),e_a,e_b\big).
\end{align}
We can split it into two parts,
\begin{align}
l_2\big(l_2(e_a, e_b),e_I\big)-2\,l_2\big(l_2(e_{[a|},e_I),e_{|b]}\big)&=f_{ab}{}^r l_2(e_r,e_I)+X_{[a|I}{}^Kl_2(e_K,e_{|b]})\nonumber\\
&=-\frac{1}{2}\Big(f_{ab}{}^rX_{r I}^J-X_{[a|I}{}^KX_{|b]K}{}^J\Big)\,e_J\label{heavycomputation0}
\end{align}
and
\begin{align}
l_1\big(l_3(e_a,e_b,e_I)\big)&+l_3\big(l_1(e_I),e_a,e_b\big)=-\frac{1}{6}\Big(f_{ab}{}^{r}b_{Itr}+X_{[a|I}{}^{K}b_{Kt|b]}\Big)\,l_1(e^t)-h_I^sl_3(e_s, e_a, e_b)\nonumber\\
&\quad=\left[-\frac{g^{Jt}}{6}\Big(f_{ab}{}^{r}b_{Itr}+X_{[a|I}{}^{K}b_{Kt|b]}\Big)+h_I^sf_{[sa}{}^rd_{b]r}^J\right]\,e_J.\label{heavycomputation}
\end{align}
One can rewrite the last term on the right hand side as
\begin{align}
h_I^sf_{[sa}{}^rd_{b]r}^J&=\frac{2}{3}h_I^sf_{s[a}{}^rd_{b]r}+\frac{1}{3}h_{I}^sf_{ab}{}^rd_{rs}^J\nonumber\\
&=\frac{2}{3}h_I^sd_{s[a|}^Kh_K^rd_{|b]r}^J+\frac{1}{6}f_{ab}{}^r(2h_I^sd^J_rs)\nonumber\\
&=\frac{1}{3}X_{[a|I}{}^Kh_K^rd_{|b]r}^J+\frac{1}{6}f_{ab}{}^r(X_{rI}{}^J+g^{Jt}b_{Itr}),
\end{align}
where we passed from the first line to the second line by using \eqref{lol5}, and from the second line from the third line by using the definition of $X_{aI}{}^K$ and by \eqref{lol4}. Then the term in the brackets on the right hand side of \eqref{heavycomputation} becomes
\begin{align}
-\frac{g^{Jt}}{6}\Big(f_{ab}{}^{r}b_{Itr}+X_{[a|I}{}^{K}b_{Kt|b]}\Big)+\frac{1}{3}X_{[a|I}{}^Kh_K^rd_{|b]r}^J&+\frac{1}{6}f_{ab}{}^r(X_{rI}{}^J+g^{Jt}b_{Itr})\nonumber\\
&=\frac{1}{6}f_{ab}{}^rX_{rI}{}^J+\frac{1}{6}X_{[a|I}{}^{K}X_{|b]K}{}^J.\label{heavycomputation2}
\end{align}
Then, adding \eqref{heavycomputation2} to \eqref{heavycomputation0}, one finds
\begin{align}
\mathrm{J}(e_a,e_b,e_I)=-\frac{1}{3}\Big(f_{ab}{}^rX_{rI}{}^J-2X_{[a|I}{}^{K}X_{|b]K}{}^J\Big)\,e_J.
\end{align}
But the right hand side is nothing but one third times
\begin{align}
\Big(\rho_{[\Theta(e_a),\Theta(e_b)]}-\big[\rho_{\Theta(e_a)},\rho_{\Theta(e_b)}\big]\Big)(e_I),
\end{align}
which vanishes because $X_1$ is an $\mathfrak{h}$-module. One can check as well that the (higher) Jacobiator of $e_a, e_b$ and $e^t$ is one third times
\begin{align}
\Big(\rho_{[\Theta(e_a),\Theta(e_b)]}-\big[\rho_{\Theta(e_a)},\rho_{\Theta(e_b)}\big]\Big)(e^t),
\end{align}
which vanishes because $X_2$ is an $\mathfrak{h}$-module.

The next higher Jacobi identity that we can compute is the one involving $e_a$ and $e_I, e_J$,
\begin{align}
\mathrm{J}(e_a,e_I,e_J)&\equiv 2\,l_2\big(l_2(e_a, e_{(I}),e_{J)}\big)+l_2\big(l_2(e_I,e_J),e_a\big)\nonumber\\
&\quad\,+l_1\big(l_3(e_a,e_I,e_J)\big)-2\,l_3\big(l_1(e_{(I|}),e_a,e_{|J)}\big).
\end{align}
Since the bracket between two elements $e_K$ and $e_L$ is always zero, the two first terms identically vanish. Now let us show that the two last terms cancel one another,
\begin{align}
l_1\big(l_3(e_a,e_I,e_J)\big)&-2\,l_3\big(l_1(e_{(I|}),e_a,e_{|J)}\big)\nonumber\\&\quad=-\frac{1}{3}X_{a(I|}{}^Kc_{\alpha K|J)}\,l_1(e^\alpha)+2h_{(I|}^{r}\,l_3(e_r,e_a,e_{|J)})\nonumber\\
&\quad=\frac{1}{3}\left[X_{a(I|}{}^Kc_{\alpha K|J)}k_t^\alpha-h_{(I|}^{r}f_{ra}{}^{s}b_{|J)ts}-h_{(I|}^{r}X_{[r|J)}{}^{K}b_{Kt|a]}\right]\,e^t.\label{heavycomputation4}
\end{align}
Let us turn our attention to the last two terms:
\begin{align}
-h_{(I|}^{r}f_{ra}{}^{s}b_{|J)ts}-h_{(I|}^{r}X_{[r|J)}{}^{K}b_{Kt|a]}&=-h_{(I|}^{r}d^K_{ra}h_K^sb_{|J)ts}+\frac{1}{2}h_{(I|}^{r}X_{a|J)}{}^{K}b_{Ktr}\nonumber\\
&=-h_{(J|}^{s}d^K_{sa}h_K^rb_{|I)tr}+\frac{1}{2}X_{a(J}{}^{K}h_{I)}^{r}b_{Ktr}\nonumber\\
&=-\frac{1}{2}X_{a(J|}{}^Kh^r_{K}b_{|I)tr}+\frac{1}{2}X_{a(J}{}^{K}h_{I)}^{r}b_{Ktr}\nonumber\\
&=X_{a(I|}{}^K\Big(-\frac{1}{2}h^r_{K}b_{|J)tr}+\frac{1}{2}h_{|J)}^{r}b_{Ktr}\Big).
\end{align}
We have passed from the left hand side of the first line to the right hand side by applying \eqref{lol5} to the first term, by developing the skew-symmetrization of $a$ and $r$, and by noticing that $h_{(I|}^rX_{r|J)}{}^K=0$ by \eqref{lol6}. The second line is just the first line, rewritten with different indices. Passing from the second line to the third line uses the definition of $X_{a(J|}{}^K$ and \eqref{lol4}. Passing to the fourth line is done by using the symmetrization between $I$ and $J$. Then, by inserting the fourth line into \eqref{heavycomputation4}, one obtains
\begin{align}
l_1\big(l_3(e_a,e_I,e_J)\big)&-2\,l_3\big(l_1(e_{(I|}),e_a,e_{|J)}\big)\nonumber\\&=\frac{1}{3}X_{a(I|}{}^K\left[k_t^\alpha c_{\alpha K|J)}-\frac{1}{2}h^r_{K}b_{|J)tr}+\frac{1}{2}h_{|J)}^{r}b_{Ktr}\right]\,e^t.
\end{align}
But the term in the bracket vanishes by \eqref{lol11}, hence proving the desired higher Jacobi identity. 

One can now compute the higher Jacobi identities mixing the 2- and 3-brackets and straightforwardly check that $l_2\circ l_3-l_3\circ l_2$ identically vanishes, which is consistent with the fact that the 4-bracket is zero. For example, for four elements $e_a, e_b, e_c, e_d$, we have (underlined indices imply full anti-symmetry on these indices):
\begin{align}
6\,l_3\big(l_2(e_{\underline{a}},e_{\underline{b}}),e_{\underline{c}},e_{\underline{d}}\big)-&4\,l_2\big(l_3(e_{\underline{a}},e_{\underline{b}},e_{\underline{c}}),e_{\underline{d}}\big)\nonumber\\
&=6f_{\underline{a}\underline{b}}{}^rl_3(e_r,e_{\underline{c}},e_{\underline{d}})+4f_{\underline{a}\underline{b}}{}^rd_{\underline{c}r}^Jl_2(e_J,e_{\underline{d}})\nonumber\\
&=\Big(-6f_{\underline{a}\underline{b}}{}^rf_{[r\underline{c}}{}^sd_{\underline{d}]s}^K+2f_{\underline{a}\underline{b}}{}^rd_{\underline{c}r}^JX_{\underline{d}J}{}^K\Big)\,e_K
\end{align}
By expanding $f_{[r\underline{c}}{}^sd_{\underline{d}]s}^K$ on the one hand, and by using the identity $X_{\underline{d}J}{}^Kd_{\underline{c}r}^J-X_{\underline{d}\underline{c}}{}^sd_{sr}^K-X_{\underline{d}r}{}^sd_{\underline{c}s}^K=0$ (which is a rewriting of \eqref{lol1}) on the other hand, the term in the parenthesis can be rewritten as:
\begin{align}
-4f_{\underline{a}\underline{b}}{}^rf_{r\underline{c}}{}^sd_{\underline{d}s}^K&\underbrace{-2f_{\underline{a}\underline{b}}{}^rf_{\underline{c}\underline{d}}{}^sd_{rs}^K+2f_{\underline{a}\underline{b}}{}^rd_{sr}^KX_{\underline{d}\underline{c}}{}^s}_{=\ 0}+2f_{\underline{a}\underline{b}}{}^rd_{\underline{c}s}^KX_{\underline{d}r}{}^s\nonumber\\
&=-4f_{\underline{a}\underline{b}}{}^rf_{r\underline{c}}{}^sd_{\underline{d}s}^K-2f_{\underline{a}\underline{b}}{}^rf_{\underline{d}r}{}^sd_{\underline{c}s}^K+2f_{\underline{a}\underline{b}}{}^rh_J^sd^J_{r\underline{d}}d_{\underline{c}s}^K\nonumber\\
&=-4f_{\underline{a}\underline{b}}{}^rf_{r\underline{c}}{}^sd_{\underline{d}s}^K+4f_{\underline{a}\underline{b}}{}^rf_{\underline{d}r}{}^sd_{\underline{c}s}^K=0,
\end{align}
 We passed from the first line to the second line by using the definition of $X_{\underline{d}r}{}^s$,
and from the second line to the last one by using \eqref{lol3}. The Jacobi identities  of the type $l_3\circ l_2-l_2\circ l_3=0$ that involve elements from $X_1$ and $X_2$ can be shown to be satisfied as well.

Now let us show that the Jacobi identity $l_3\circ l_3+l_1\circ l_5=0$ is satisfied on $X_0$. For five elements $e_a, e_b, e_c, e_d, e_e$, we have (underlined indices imply full anti-symmetry on these indices):
\begin{align}
10\,l_3\big(l_3(e_{\underline{a}},&e_{\underline{b}},e_{\underline{c}}),e_{\underline{d}},e_{\underline{e}}\big)+l_1\big(l_5(e_{\underline{a}},e_{\underline{b}},e_{\underline{c}},e_{\underline{d}},e_{\underline{e}})\big)\nonumber\\
&=-10f_{\underline{a}\underline{b}}^rd_{\underline{c}r}^Il_3(e_I,e_{\underline{d}},e_{\underline{e}})+\frac{1}{3}f_{\underline{a}\underline{b}}{}^rd_{\underline{c}r}^Ib_{Is\underline{d}}c_{\alpha\underline{e}}^s\,l_1(e^\alpha)\\
&=\left(\frac{5}{3}f_{\underline{a}\underline{b}}^rd_{\underline{c}r}^I\big(f_{\underline{d}\underline{e}}{}^{s}b_{Its}+X_{\underline{d}I}{}^Kb_{Kt\underline{e}}\big)-\frac{1}{3}f_{\underline{a}\underline{b}}{}^rd_{\underline{c}r}^Ib_{Is\underline{d}}c_{\alpha\underline{e}}^sk^\alpha_t\right)e^t
\end{align}
By using \eqref{lol12} and by noticing that $f_{\underline{d}\underline{e}}{}^s=X_{\underline{e}\underline{d}}{}^s$, the term in parenthesis becomes one third
\begin{align}
f_{\underline{a}\underline{b}}^rd_{\underline{c}r}^I\Big(5X_{\underline{e}\underline{d}}{}^sb_{Its}&-5X_{\underline{e}I}{}^Kb_{Kt\underline{d}}-b_{Is\underline{d}}X_{\underline{e}t}{}^s+b_{Is\underline{d}}g^{Ks}b_{Kt\underline{e}}\Big)\nonumber\\
&=f_{\underline{a}\underline{b}}^rd_{\underline{c}r}^I\Big(6X_{\underline{e}\underline{d}}{}^sb_{Its}-4X_{\underline{e}I}{}^Kb_{Kt\underline{d}}-b_{Is\underline{e}}g^{Ks}b_{Kt\underline{d}}\Big)\nonumber\\
&=f_{\underline{a}\underline{b}}^rd_{\underline{c}r}^I\Big(6X_{\underline{e}\underline{d}}{}^sb_{Its}-3X_{\underline{e}I}{}^Kb_{Kt\underline{d}}-2h_I^sd^K_{\underline{e}s}b_{Kt\underline{d}}\Big)\nonumber\\
&=f_{\underline{a}\underline{b}}^r\Big(6X_{\underline{e}\underline{d}}{}^sd_{\underline{c}r}^Ib_{Its}-3X_{\underline{e}I}{}^Kd_{\underline{c}r}^Ib_{Kt\underline{d}}+3X_{\underline{e}r}{}^sd_{\underline{c}s}^Kb_{Kt\underline{d}}\Big)\nonumber\\
&=-3X_{\underline{a}\underline{b}}^r\Big(2X_{\underline{e}\underline{d}}{}^sd_{\underline{c}r}^Kb_{Kts}-X_{\underline{e}\underline{c}}{}^sd_{rs}^Kb_{Kt\underline{d}}\Big)\nonumber\\
&=-3\Big(X_{\underline{a}\underline{b}}^rX_{\underline{e}\underline{d}}{}^sd_{\underline{c}r}^Kb_{Kts}+X_{\underline{a}\underline{b}}^sX_{\underline{e}\underline{d}}{}^rd_{\underline{c}s}^Kb_{Ktr}+X_{\underline{a}\underline{b}}^rX_{\underline{e}\underline{d}}{}^sd_{rs}^Kb_{Kt\underline{c}}\Big)\nonumber\\
&=-9X_{\underline{a}\underline{b}}^rX_{\underline{e}\underline{d}}{}^sb_{Kt(s}d^K_{\underline{c}r)}=0
\end{align}
We passed from the first line to second line by using the identity $-X_{\underline{e}I}{}^Kb_{Kt\underline{d}}-b_{Is\underline{d}}X_{\underline{e}t}{}^s=X_{\underline{e}\underline{d}}{}^sb_{Its}$ which is a rewriting of \eqref{lol2}, from the second line to the third by using the definition of $X_{\underline{e}I}{}^K$, and from the third line to the fourth by noticing that $2f_{\underline{a}\underline{b}}^rd_{\underline{c}r}^Ih_I^s=3f_{\underline{a}\underline{b}}^rX_{\underline{c}r}{}^s$, which follows from \eqref{lol3}. We passed from the fourth line to the fifth one 
 by using the identity $-X_{\underline{e}I}{}^Kd_{\underline{c}r}^I+X_{\underline{e}r}{}^sd_{\underline{c}s}^K=-X_{\underline{e}\underline{c}}{}^sd_{rs}^K$ which is a rewriting of \eqref{lol1}. From the fifth line to the sixth line we have just rewritten the first term in the parenthesis, and the symmetry properties of the three terms give the last line, which vanishes by \eqref{lol8}.

We have thus proven that the brackets \eqref{eq:infinity1}--\eqref{eq:infinity3} satisfy the higher Jacobi identities \eqref{higherJac}, hence defining a $L_\infty$-algebra. Note that this $L_\infty$-algebra associated to the $(1,0)$ superconformal model does not coincide with the one found in \cite{Lavau:2014iva}, which was obtained from the Bianchi identities, although the
underlying 
tensor hierarchy is the same. We leave for future work a more general study on the relationship between those two $L_\infty$-algebras.

\section*{Acknowledgments}

We would like to thank Roberto Bonezzi, Martin Cederwall, Olaf Hohm and Jim Stasheff for for discussions and comments on the first version of this paper.
In particular we are grateful to Olaf Hohm for explaining the ideas behind the work
\cite{Bonezzi:2019ygf}. We would also like to thank the anonymous referee for 
suggesting some well justified clarifications on the role of the degree-zero subspace of the differential graded Lie algebras.
This work was initiated during a visit at Institut des Hautes \'Etudes Scientifiques (IH\'ES), and we would like to thank the institute for its hospitality.
The work of JP is supported by the Swedish Research Council, project no.\ 2015-02468. The work of SL is supported by the Agence Nationale de la Recherche, project SINGSTAR.

\phantomsection
%
%\addcontentsline{toc}{section}{References}
\bibliographystyle{utphysmod2}

%\bibliography{biblio.bib}

%\bibliography{biblio}

\begin{thebibliography}{10}

\bibitem{Bonezzi:2019ygf}
R.~Bonezzi and O.~Hohm,  {\em {Leibniz Gauge Theories and Infinity
  Structures}}, Commun. Math. Phys., \textbf{377}, 2027--2077 (2020)
[\href{http://www.arXiv.org/abs/1904.11036}{{\tt 1904.11036}}].
%%CITATION = ARXIV:1904.11036;%%.

\bibitem{Loday}
J.-L. Loday and T.~Pirashvili,  {\em Universal enveloping algebras of Leibniz
  algebras and (co)homology}, Math.\ Annal. {\bf 296}, 139--158 (1993).

\bibitem{deWit:2002vt}
B.~de~Wit, H.~Samtleben and M.~Trigiante,  {\em {On Lagrangians and gaugings of
  maximal supergravities}}, Nucl.\ Phys.\ {\bf B655}, 93--126 (2003)
[\href{http://www.arXiv.org/abs/hep-th/0212239}{{\tt hep-th/0212239}}].
%%CITATION = HEP-TH/0212239;%%.

\bibitem{deWit:2004nw}
B.~de~Wit, H.~Samtleben and M.~Trigiante,  {\em {The Maximal $D=5$
  supergravities}}, Nucl.\ Phys.\ {\bf B716}, 215--247 (2005)
[\href{http://www.arXiv.org/abs/hep-th/0412173}{{\tt hep-th/0412173}}].
%%CITATION = HEP-TH/0412173;%%.

\bibitem{deWit:2005hv}
B.~de~Wit and H.~Samtleben,  {\em {Gauged maximal supergravities and
  hierarchies of nonAbelian vector-tensor systems}}, Fortsch.\ Phys.\ {\bf 53},
  442--449 (2005)
[\href{http://www.arXiv.org/abs/hep-th/0501243}{{\tt hep-th/0501243}}].
%%CITATION = HEP-TH/0501243;%%.

\bibitem{deWit:2008ta}
B.~de~Wit, H.~Nicolai and H.~Samtleben,  {\em {Gauged supergravities, tensor
  hierarchies, and M-theory}}, JHEP {\bf 0802}, 044 (2008)
[\href{http://www.arXiv.org/abs/0801.1294}{{\tt 0801.1294}}].
%%CITATION = ARXIV:0801.1294;%%.

\bibitem{Trigiante:2016mnt}
M.~Trigiante,  {\em {Gauged Supergravities}}, Phys.\ Rept.\ {\bf 680}, 1--175
  (2017)
[\href{http://www.arXiv.org/abs/1609.09745}{{\tt 1609.09745}}].
%%CITATION = ARXIV:1609.09745;%%.

\bibitem{Hull:2009zb}
C.~Hull and B.~Zwiebach,  {\em {The Gauge algebra of double field theory and
  Courant brackets}}, JHEP {\bf 09}, 090 (2009)
[\href{http://www.arXiv.org/abs/0908.1792}{{\tt 0908.1792}}].
%%CITATION = ARXIV:0908.1792;%%.

\bibitem{Coimbra:2011ky}
A.~Coimbra, C.~Strickland-Constable and D.~Waldram,  {\em {$E_{d(d)} \times
  \mathbb{R}^+$ generalised geometry, connections and M-theory}}, JHEP {\bf
  1402}, 054 (2014)
[\href{http://www.arXiv.org/abs/1112.3989}{{\tt 1112.3989}}].
%%CITATION = ARXIV:1112.3989;%%.

\bibitem{Berman:2012vc}
D.~S. Berman, M.~Cederwall, A.~Kleinschmidt and D.~C. Thompson,  {\em {The
  gauge structure of generalised diffeomorphisms}}, JHEP {\bf 01}, 064 (2013)
[\href{http://www.arXiv.org/abs/1208.5884}{{\tt 1208.5884}}].
%%CITATION = ARXIV:1208.5884;%%.

\bibitem{Cederwall:2013naa}
M.~Cederwall, J.~Edlund and A.~Karlsson,  {\em {Exceptional geometry and tensor
  fields}}, JHEP {\bf 07}, 028 (2013)
[\href{http://www.arXiv.org/abs/1302.6736}{{\tt 1302.6736}}].
%%CITATION = ARXIV:1302.6736;%%.

\bibitem{Cederwall:2013oaa}
M.~Cederwall,  {\em {Non-gravitational exceptional supermultiplets}}, JHEP {\bf
  07}, 025 (2013)
[\href{http://www.arXiv.org/abs/1302.6737}{{\tt 1302.6737}}].
%%CITATION = ARXIV:1302.6737;%%.

\bibitem{Aldazabal:2013mya}
G.~Aldazabal, M.~Gra\~{n}a, D.~Marqu\'es and J.~Rosabal,  {\em {Extended
  geometry and gauged maximal supergravity}}, JHEP {\bf 1306}, 046 (2013)
[\href{http://www.arXiv.org/abs/1302.5419}{{\tt 1302.5419}}].
%%CITATION = ARXIV:1302.5419;%%.

\bibitem{Hohm:2013nja}
O.~Hohm and H.~Samtleben,  {\em {Gauge theory of Kaluza-Klein and winding
  modes}}, Phys.\ Rev.\ {\bf D88}, 085005 (2013)
[\href{http://www.arXiv.org/abs/1307.0039}{{\tt 1307.0039}}].
%%CITATION = ARXIV:1307.0039;%%.

\bibitem{Hohm:2013pua}
O.~Hohm and H.~Samtleben,  {\em Exceptional form of ${D}=11$ supergravity},
  Phys.\ Rev.\ Lett.\ {\bf 111}, 231601 (2013)
[\href{http://www.arXiv.org/abs/1308.1673}{{\tt 1308.1673}}].
%%CITATION = ARXIV:1308.1673;%%.

\bibitem{Hohm:2013vpa}
O.~Hohm and H.~Samtleben,  {\em Exceptional field theory {I}: {E}$_{6(6)}$
  covariant form of {M}-theory and type {IIB}}, Phys.\ Rev.\ {\bf D89}, 066016
  (2014)
[\href{http://www.arXiv.org/abs/1312.0614}{{\tt 1312.0614}}].
%%CITATION = ARXIV:1312.0614;%%.

\bibitem{Hohm:2013uia}
O.~Hohm and H.~Samtleben,  {\em Exceptional field theory {II}: {E}$_{7(7)}$},
  Phys.\ Rev.\ {\bf D89}, 066017 (2014)
[\href{http://www.arXiv.org/abs/1312.4542}{{\tt 1312.4542}}].
%%CITATION = ARXIV:1312.4542;%%.

\bibitem{Hohm:2014fxa}
O.~Hohm and H.~Samtleben,  {\em {Exceptional field theory. III. E$_{8(8)}$}},
  Phys.\ Rev.\ {\bf D90}, 066002 (2014)
[\href{http://www.arXiv.org/abs/1406.3348}{{\tt 1406.3348}}].
%%CITATION = ARXIV:1406.3348;%%.

\bibitem{Hohm:2015xna}
O.~Hohm and Y.-N. Wang,  {\em {Tensor hierarchy and generalized Cartan calculus
  in $SL(3) \times SL(2)$ exceptional field theory}}, JHEP {\bf 04}, 050 (2015)
[\href{http://www.arXiv.org/abs/1501.01600}{{\tt 1501.01600}}].
%%CITATION = ARXIV:1501.01600;%%.

\bibitem{Abzalov:2015ega}
A.~Abzalov, I.~Bakhmatov and E.~T. Musaev,  {\em {Exceptional field theory:
  $SO(5,5)$}}, JHEP {\bf 06}, 088 (2015)
[\href{http://www.arXiv.org/abs/1504.01523}{{\tt 1504.01523}}].
%%CITATION = ARXIV:1504.01523;%%.

\bibitem{Wang:2015hca}
Y.-N. Wang,  {\em {Generalized Cartan calculus in general dimension}}, JHEP
  {\bf 07}, 114 (2015)
[\href{http://www.arXiv.org/abs/1504.04780}{{\tt 1504.04780}}].
%%CITATION = ARXIV:1504.04780;%%.

\bibitem{Cederwall:2015ica}
M.~Cederwall and J.~A. Rosabal,  {\em {E$_{8}$ geometry}}, JHEP {\bf 07}, 007
  (2015)
[\href{http://www.arXiv.org/abs/1504.04843}{{\tt 1504.04843}}].
%%CITATION = ARXIV:1504.04843;%%.

\bibitem{Musaev:2015ces}
E.~T. Musaev,  {\em {Exceptional field theory: $SL(5)$}}, JHEP {\bf 02}, 012
  (2016)
[\href{http://www.arXiv.org/abs/1512.02163}{{\tt 1512.02163}}].
%%CITATION = ARXIV:1512.02163;%%.

\bibitem{Berman:2015rcc}
D.~S. Berman, C.~D.~A. Blair, E.~Malek and F.~J. Rudolph,  {\em {An action for
  F-theory: $\mathrm{SL}(2)\times {{\mathbb{R}}}^{+}$ exceptional field theory}},
  Class. Quant. Grav. {\bf 33}, 195009 (2016)
[\href{http://www.arXiv.org/abs/1512.06115}{{\tt 1512.06115}}].
%%CITATION = ARXIV:1512.06115;%%.

\bibitem{Deser:2016qkw}
A.~Deser and C.~Saemann,  {\em {Extended Riemannian Geometry I: Local Double
  Field Theory}}, C.\ Ann.\ Henri Poincar\'e {\bf 19, 2297} (2018)
[\href{http://www.arXiv.org/abs/1611.02772}{{\tt 1611.02772}}].
%%CITATION = ARXIV:1611.02772;%%.

\bibitem{Bossard:2017aae}
G.~Bossard, M.~Cederwall, A.~Kleinschmidt, J.~Palmkvist and H.~Samtleben,  {\em
  {Generalized diffeomorphisms for $E_9$}}, Phys. Rev. {\bf D96}, 106022 (2017)
[\href{http://www.arXiv.org/abs/1708.08936}{{\tt 1708.08936}}].
%%CITATION = ARXIV:1708.08936;%%.

\bibitem{Cederwall:2017fjm}
M.~Cederwall and J.~Palmkvist,  {\em {Extended geometries}}, JHEP {\bf 02}, 071
  (2018)
[\href{http://www.arXiv.org/abs/1711.07694}{{\tt 1711.07694}}].
%%CITATION = ARXIV:1711.07694;%%.

\bibitem{Cagnacci2019}
Y.~Cagnacci, T.~Codina and D.~Marques,  {\em $L_\infty$ algebras and tensor
  hierarchies in exceptional field theory and gauged supergravity}, JHEP {\bf
  2019}, 117 (2019) [\href{http://www.arXiv.org/abs/1807.06028}{{\tt
  1807.06028}}].

\bibitem{Cederwall:2018aab}
M.~Cederwall and J.~Palmkvist,  {\em {$L_\infty$ algebras for extended geometry
  from Borcherds superalgebras}}, Commun.\ Math.\ Phys.\  {\bf 369}, 721 (2019)
[\href{http://www.arXiv.org/abs/1804.04377}{{\tt 1804.04377}}].
%%CITATION = ARXIV:1804.04377;%%.

\bibitem{Hohm:2018ybo}
O.~Hohm and H.~Samtleben,  {\em {Leibniz-Chern-Simons Theory and Phases of
  Exceptional Field Theory}}, Commun. Math. Phys. {\bf 369}, 1055 (2019)
[\href{http://www.arXiv.org/abs/1805.03220}{{\tt 1805.03220}}].
%%CITATION = ARXIV:1805.03220;%%.

\bibitem{Bossard:2018utw}
G.~Bossard, F.~Ciceri, G.~Inverso, A.~Kleinschmidt and H.~Samtleben,  {\em
  {E$_{9}$ exceptional field theory. Part I. The potential}}, JHEP {\bf 03},
  089 (2019)
[\href{http://www.arXiv.org/abs/1811.04088}{{\tt 1811.04088}}].
%%CITATION = ARXIV:1811.04088;%%.

\bibitem{Hohm:2019wql}
O.~Hohm and H.~Samtleben,  {\em {Higher Gauge Structures in Double and
  Exceptional Field Theory}}, in {\em {Durham Symposium, Higher Structures in
  M-Theory Durham, UK, August 12-18, 2018}}.
\newblock 2019.
\newblock
[\href{http://www.arXiv.org/abs/1903.02821}{{\tt 1903.02821}}].
\newblock
%%CITATION = ARXIV:1903.02821;%%.

\bibitem{Strobl:2016aph}
T.~Strobl,  {\em {Non-abelian Gerbes and Enhanced Leibniz Algebras}}, Phys.
  Rev. {\bf D94}, 021702 (2016)
[\href{http://www.arXiv.org/abs/1607.00060}{{\tt 1607.00060}}].
%%CITATION = ARXIV:1607.00060;%%.

\bibitem{Strobl:2019hha}
T.~Strobl and F.~Wagemann,  {\em {Enhanced Leibniz Algebras: Structure Theorem
  and Induced Lie 2-Algebra}}
[\href{http://www.arXiv.org/abs/1901.01014}{{\tt 1901.01014}}].
%%CITATION = ARXIV:1901.01014;%%.

\bibitem{Palmkvist:2013vya}
J.~Palmkvist,  {\em {The tensor hierarchy algebra}}, J. Math. Phys. {\bf 55},
  011701 (2014)
[\href{http://www.arXiv.org/abs/1305.0018}{{\tt 1305.0018}}].
%%CITATION = ARXIV:1305.0018;%%.

\bibitem{Greitz:2013pua}
J.~Greitz, P.~Howe and J.~Palmkvist,  {\em {The tensor hierarchy simplified}},
Class.\ Quant.\ Grav.\  {\bf 31}, 087001 (2014)
[\href{http://www.arXiv.org/abs/1308.4972}{{\tt 1308.4972}}].
%%CITATION = ARXIV:1308.4972;%%.

\bibitem{Henneaux:2010ys}
M.~Henneaux, B.~L. Julia and J.~Levie,  {\em {$E_{11}$, Borcherds algebras and
  maximal supergravity}}, JHEP {\bf 1204}, 078 (2012)
[\href{http://www.arXiv.org/abs/1007.5241}{{\tt 1007.5241}}].
%%CITATION = ARXIV:1007.5241;%%.

\bibitem{Palmkvist:2011vz}
J.~Palmkvist,  {\em {Tensor hierarchies, Borcherds algebras and $E_{11}$}}, JHEP
  {\bf 1202}, 066 (2012)
[\href{http://www.arXiv.org/abs/1110.4892}{{\tt 1110.4892}}].
%%CITATION = ARXIV:1110.4892;%%.

\bibitem{Cederwall:2015oua}
M.~Cederwall and J.~Palmkvist,  {\em {Superalgebras, constraints and partition
  functions}}, JHEP {\bf 08}, 036 (2015)
[\href{http://www.arXiv.org/abs/1503.06215}{{\tt 1503.06215}}].
%%CITATION = ARXIV:1503.06215;%%.

\bibitem{Palmkvist:2015dea}
J.~Palmkvist,  {\em {Exceptional geometry and Borcherds superalgebras}}, JHEP
  {\bf 11}, 032 (2015)
[\href{http://www.arXiv.org/abs/1507.08828}{{\tt 1507.08828}}].
%%CITATION = ARXIV:1507.08828;%%.

\bibitem{Lavau:2017tvi}
S.~Lavau,  {\em {Tensor hierarchies and Leibniz algebras}}, J. Geom. Phys. {\bf
  144}, 147--189 (2019)
[\href{http://www.arXiv.org/abs/1708.07068}{{\tt 1708.07068}}].
%%CITATION = ARXIV:1708.07068;%%.

\bibitem{Lada:1992wc}
T.~Lada and J.~Stasheff,  {\em {Introduction to SH Lie algebras for
  physicists}}, Int. J. Theor. Phys. {\bf 32}, 1087--1104 (1993)
[\href{http://www.arXiv.org/abs/hep-th/9209099}{{\tt hep-th/9209099}}].
%%CITATION = HEP-TH/9209099;%%.

\bibitem{Marklada}
T.~Lada and M.~Markl,  {\em {Strongly homotopy Lie algebras}}, Commun. Algebr.
  {\bf 23}, 2147--2161 (1994).

\bibitem{Palmer:2013pka}
S.~Palmer and C.~Saemann,  {\em {Six-Dimensional (1,0) Superconformal Models
  and Higher Gauge Theory}}, J. Math. Phys. {\bf 54}, 113509 (2013)
[\href{http://www.arXiv.org/abs/1308.2622}{{\tt 1308.2622}}].
%%CITATION = ARXIV:1308.2622;%%.

\bibitem{Lavau:2014iva}
S.~Lavau, H.~Samtleben and T.~Strobl,  {\em {Hidden Q-structure and Lie
  3-algebra for non-abelian superconformal models in six dimensions}}, J. Geom.
  Phys. {\bf 86}, 497--533 (2014)
[\href{http://www.arXiv.org/abs/1403.7114}{{\tt 1403.7114}}].
%%CITATION = ARXIV:1403.7114;%%.

\bibitem{Ritter:2015ymv}
P.~Ritter and C.~Saemann,  {\em {$L_\infty$-Algebra Models and Higher
  Chern-Simons Theories}}, Rev. Math. Phys. {\bf 28}, 1650021 (2016)
[\href{http://www.arXiv.org/abs/1511.08201}{{\tt 1511.08201}}].
%%CITATION = ARXIV:1511.08201;%%.

\bibitem{Saemann:2017rjm}
C.~Saemann and L.~Schmidt,  {\em {The Non-Abelian Self-Dual String and the
  (2,0)-Theory}}
[\href{http://www.arXiv.org/abs/1705.02353}{{\tt 1705.02353}}].
%%CITATION = ARXIV:1705.02353;%%.

\bibitem{Hohm:2017pnh}
O.~Hohm and B.~Zwiebach,  {\em {$L_{\infty}$ algebras and field theory}},
  Fortsch. Phys. {\bf 65}, 1700014 (2017)
[\href{http://www.arXiv.org/abs/1701.08824}{{\tt 1701.08824}}].
%%CITATION = ARXIV:1701.08824;%%.

\bibitem{Deser:2018oyg}
A.~Deser and C.~Saemann,  {\em {Derived brackets and symmetries in generalized
  geometry and double field theory}}, in {\em {17th Hellenic School and
  Workshops on Elementary Particle Physics and Gravity (CORFU2017) Corfu,
  Greece, September 2-28, 2017}}.
\newblock 2018.
\newblock
[\href{http://www.arXiv.org/abs/1803.01659}{{\tt 1803.01659}}].
\newblock
%%CITATION = ARXIV:1803.01659;%%.

\bibitem{Jurco:2019woz}
B.~Jur$\rm\check{c}$o, C.~Saemann, U.~Schreiber and M.~Wolf,  {\em {Higher
  Structures in M-Theory}}, in {\em {Durham Symposium, Higher Structures in
  M-Theory Durham, UK, August 12-18, 2018}}.
\newblock 2019.
\newblock
[\href{http://www.arXiv.org/abs/1903.02807}{{\tt 1903.02807}}].
\newblock
%%CITATION = ARXIV:1903.02807;%%.

\bibitem{Getzler:2010}
E.~Getzler,  {\em {Higher derived brackets}}
  [\href{http://www.arXiv.org/abs/1010.5859}{{\tt 1010.5859}}].

\bibitem{FiorenzaManetti}
D.~Fiorenza and M.~Manetti,  {\em {L-infinity structures on mapping cones}},
  Algebra \& Number Theory {\bf 1}, 301--330 (2007)
[\href{http://www.arXiv.org/abs/math/0601312}{{\tt math/0601312}}].
%%CITATION = ARXIV:math/0601312;%%.

\bibitem{Kotov:2018vcz}
A.~Kotov and T.~Strobl,  {\em {The Embedding Tensor, Leibniz-Loday Algebras,
  and Their Higher Gauge Theories}}
[\href{http://www.arXiv.org/abs/1812.08611}{{\tt 1812.08611}}].
%%CITATION = ARXIV:1812.08611;%%.

\bibitem{Samtleben:2011fj}
H.~Samtleben, E.~Sezgin and R.~Wimmer,  {\em {(1,0) superconformal models in
  six dimensions}}, JHEP {\bf 12}, 062 (2011)
[\href{http://www.arXiv.org/abs/1108.4060}{{\tt 1108.4060}}].
%%CITATION = ARXIV:1108.4060;%%.

\bibitem{Carbone:2018njd}
L.~Carbone, M.~Cederwall and J.~Palmkvist,  {\em {Generators and relations for
  Lie superalgebras of Cartan type}}
[\href{http://www.arXiv.org/abs/1802.05767}{{\tt 1802.05767}}].
%%CITATION = ARXIV:1802.05767;%%.

\bibitem{Bossard:2017wxl}
G.~Bossard, A.~Kleinschmidt, J.~Palmkvist, C.~N. Pope and E.~Sezgin,  {\em
  {Beyond $E_{11}$}}, JHEP {\bf 05}, 020 (2017)
[\href{http://www.arXiv.org/abs/1703.01305}{{\tt 1703.01305}}].
%%CITATION = ARXIV:1703.01305;%%.

\bibitem{Bossard:2019ksx}
G.~Bossard, A.~Kleinschmidt and E.~Sezgin,  {\em {On supersymmetric $E_{11}$
  exceptional field theory}},
[\href{http://www.arXiv.org/abs/1907.02080}{{\tt 1907.02080}}].
%%CITATION = ARXIV:1907.02080;%%.

\bibitem{CederwallPalmkvistTHA}
M.~Cederwall and J.~Palmkvist,  {\em {Extended geometry and tensor hierarchy
  algebras}}, to appear.
%\href{http://www.arXiv.org/abs/yymm.nnnnn}{{\tt yymm.nnnnn}}].
%%CITATION = ARXIV:1804.04377;%%.

\bibitem{Kosmann:2003}
Y.~{Kosmann-Schwarzbach},  {\em {Derived Brackets}}, Lett.\ in Math.\
  Phys.\ {\bf 69}, 61--87 (2004)
  [\href{http://www.arXiv.org/abs/math/0312524}{{\tt math/0312524}}].

\bibitem{Kantor-graded}
I.~L. Kantor,  {\em Graded {L}ie algebras}, Trudy Sem.\ Vect.\ Tens.\ Anal.\ {\bf
  15}, 227--266 (1970).

\bibitem{Palmkvist:2009qq}
J.~Palmkvist,  {\em {Three-algebras, triple systems and 3-graded Lie
  superalgebras}}, J.\ Phys.\ A {\bf A43}, 015205 (2010)
[\href{http://www.arXiv.org/abs/0905.2468}{{\tt 0905.2468}}].
%%CITATION = ARXIV:0905.2468;%%.

\bibitem{Mehta}
R.~{Mehta} and M.~{Zambon},  {\em {L-infinity algebra actions}}, Differ.\ Geom.\
  Appl.\ {\bf 30}, 576--587 (2012)
  [\href{http://www.arXiv.org/abs/1202.2607}{{\tt 1202.2607}}].

\bibitem{Bering:2006eb}
K.~Bering,  {\em {On non-commutative Batalin-Vilkovisky algebras, strongly
  homotopy Lie algebras and the Courant bracket}}, Commun.\ Math.\ Phys.\ {\bf
  274}, 297--341 (2007)
[\href{http://www.arXiv.org/abs/hep-th/0603116}{{\tt hep-th/0603116}}].
%%CITATION = HEP-TH/0603116;%%.

\end{thebibliography}

\providecommand{\href}[2]{#2}\begingroup\raggedright\endgroup

\end{document}